\documentclass[12pt]{article}
\usepackage{amsmath} 
\usepackage{graphicx}
\usepackage{cite} 
\usepackage{hyperref}
\usepackage{natbib}
\usepackage{float}
\usepackage{geometry}
\usepackage{pgfplots}
\usepackage[framemethod=TikZ]{mdframed}
\usepackage{enumitem}
\usepackage{lipsum}
\usepackage{subcaption}
\usepackage[most]{tcolorbox}
\usepackage{setspace}
\usepackage{textcomp}
\usepackage{tabularx}
\usepackage{array}
\usepackage{tikz}
\usepackage{multirow}
\usetikzlibrary{arrows,positioning, automata,arrows.meta, shapes.geometric, decorations.pathreplacing}

\newcommand{\MeanDeal}{0.5}
\newcommand{\VarDeal}{0.25}

\newcommand{\MeanFinalPrice}{186.53}
\newcommand{\VarFinalPrice}{3879.23}

\newcommand{\MeanBail}{54428.57}

\newcommand{\MeanHire}{0.62}
\newcommand{\VarHire}{0.24}

\newcommand{\MeanMagPred}{13.2}
\newcommand{\MeanMagNoOutlier}{5.3}
\newcommand{\MSEwSCM}{1505}
\newcommand{\MSEwoSCM}{8628}
\newcommand{\MSEwoSCMTheory}{8915}
\newcommand{\MSEwSCMTheory}{1761}

\newcommand{\MSETheory}{128}
\newcommand{\MSEMechanistic}{725}

\newcommand{\SCMTheorySimpleRsq}{0.977}
\newcommand{\SCMTheorySimpleMean}{180.99}
\newcommand{\SCMTheorySimpleVar}{4565.99}
\newcommand{\SCMTheorySimpleEst}{0.912}
\newcommand{\SCMTheorySimpleSE}{0.009}

\newcommand{\SCMTheoryFullEst}{0.826}
\newcommand{\SCMTheoryFullSE}{0.018}

\newcommand{\SCMTheoryBidOneEst}{0.047}
\newcommand{\SCMTheoryBidOneSE}{0.009}

\newcommand{\SCMTheoryBidTwoEst}{0.039}
\newcommand{\SCMTheoryBidTwoSE}{0.008}

\newcommand{\SCMTheoryBidThreeEst}{0.03}
\newcommand{\SCMTheoryBidThreeSE}{0.009}

\newcommand{\BudgetPred}{0.05}
\newcommand{\BudgetPredMean}{0.117}
\newcommand{\BudgetPredSE}{0.016}
\newcommand{\BudgetPredSig}{*}
\newcommand{\BudgetPredSigMean}{*}
\newcommand{\MinPricePred}{-0.07}
\newcommand{\MinPricePredMean}{0.008}
\newcommand{\MinPricePredSE}{0.018}
\newcommand{\MinPricePredSig}{*}
\newcommand{\MinPricePredSigMean}{*}
\newcommand{\LovePred}{0.02}
\newcommand{\LovePredMean}{0.062}
\newcommand{\LovePredSE}{0.013}
\newcommand{\LovePredSig}{\phantom{{*}}}
\newcommand{\LovePredSigMean}{\phantom{{*}}}
\newcommand{\BidOnePred}{0.5}
\newcommand{\BidOnePredMean}{1.279}
\newcommand{\BidOnePredSE}{0.501}
\newcommand{\BidOnePredSig}{*}
\newcommand{\BidOnePredSigMean}{*}
\newcommand{\BidTwoPred}{0.5}
\newcommand{\BidTwoPredMean}{1.263}
\newcommand{\BidTwoPredSE}{0.501}
\newcommand{\BidTwoPredSig}{*}
\newcommand{\BidTwoPredSigMean}{*}
\newcommand{\BidThreePred}{0.5}
\newcommand{\BidThreePredMean}{1.269}
\newcommand{\BidThreePredSE}{0.501}
\newcommand{\BidThreePredSig}{*}
\newcommand{\BidThreePredSigMean}{*}
\newcommand{\ConvictPred}{5000}
\newcommand{\ConvictPredMean}{1785.192}
\newcommand{\ConvictPredSE}{157.347}
\newcommand{\ConvictPredSig}{*}
\newcommand{\ConvictPredSigMean}{*}
\newcommand{\CasesPred}{-200}
\newcommand{\CasesPredMean}{644.316}
\newcommand{\CasesPredSE}{79.919}
\newcommand{\CasesPredSig}{\phantom{{*}}}
\newcommand{\CasesPredSigMean}{*}
\newcommand{\RemorsePred}{-3000}
\newcommand{\RemorsePredMean}{-879.945}
\newcommand{\RemorsePredSE}{92.7}
\newcommand{\RemorsePredSig}{*}
\newcommand{\RemorsePredSigMean}{*}
\newcommand{\BarPred}{0.6}
\newcommand{\BarPredMean}{0.408}
\newcommand{\BarPredSE}{0.018}
\newcommand{\BarPredSig}{*}
\newcommand{\BarPredSigMean}{*}
\newcommand{\HeightPred}{0.1}
\newcommand{\HeightPredMean}{0.108}
\newcommand{\HeightPredSE}{0.009}
\newcommand{\HeightPredSig}{\phantom{{*}}}
\newcommand{\HeightPredSigMean}{\phantom{{*}}}
\newcommand{\FriendPred}{0.2}
\newcommand{\FriendPredMean}{0.236}
\newcommand{\FriendPredSE}{0.015}
\newcommand{\FriendPredSig}{\phantom{{*}}}
\newcommand{\FriendPredSigMean}{*}

\newcommand{\TrueBudgetBeta}{-0.111}
\newcommand{\TrueBudgetSe}{0.031}

\newcommand{\TrueBudgetPvalString}{p<0.001}
\newcommand{\TrueSellerMinBeta}{0.069}
\newcommand{\TrueSellerMinSe}{0.031}

\newcommand{\TrueSellerMinPvalString}{p=0.026}
\newcommand{\TrueLoveBeta}{0.222}
\newcommand{\TrueLoveSe}{0.153}

\newcommand{\TrueLovePvalString}{p=0.147}
\newcommand{\MissBudgetBeta}{-0.051}
\newcommand{\MissBudgetSe}{0.039}

\newcommand{\MissBudgetPvalString}{p=0.189}
\newcommand{\MissSellerMinBeta}{0.012}
\newcommand{\MissSellerMinSe}{0.037}

\newcommand{\MissSellerMinPvalString}{p=0.755}
\newcommand{\MissLoveBeta}{0.182}
\newcommand{\MissLoveSe}{0.153}

\newcommand{\MissDealBeta}{-1.622}
\newcommand{\MissDealSe}{0.615}

\newcommand{\MissDealPvalString}{p=0.008}

\newcommand{\BetaBudget}{0.037}
\newcommand{\cdBudget}{0.51}
\newcommand{\BudgetPvalString}{p<0.001}
\newcommand{\SEBudget}{0.003}

\newcommand{\BudgetSig}{*}
\newcommand{\BudgetTTestPval}{p<0.001}
\newcommand{\BudgetTTestPvalMean}{p<0.001}
\newcommand{\BudgetSign}{Yes}
\newcommand{\BudgetSignMean}{Yes}
\newcommand{\BudgetMag}{1.35}
\newcommand{\BudgetMagMean}{3.16}
\newcommand{\BetaWrittenBudget}{3.7}

\newcommand{\BetaMinPrice}{-0.035}
\newcommand{\cdMinPrice}{-0.49}
\newcommand{\MinPricePvalString}{p<0.001}
\newcommand{\SEMinPrice}{0.002}

\newcommand{\MinPriceSig}{*}
\newcommand{\MinPriceTTestPval}{p<0.001}
\newcommand{\MinPriceTTestPvalMean}{p=0.019}
\newcommand{\MinPriceSign}{Yes}
\newcommand{\MinPriceSignMean}{No}
\newcommand{\MinPriceMag}{2}
\newcommand{\MinPriceMagMean}{0.23}
\newcommand{\BetaWrittenMinPrice}{3.5}

\newcommand{\BetaLove}{-0.025}
\newcommand{\cdLove}{-0.07}
\newcommand{\LovePvalString}{p=0.044}
\newcommand{\SELove}{0.012}

\newcommand{\LoveSig}{*}
\newcommand{\LoveTTestPval}{p<0.001}
\newcommand{\LoveTTestPvalMean}{p<0.001}
\newcommand{\LoveSign}{No}
\newcommand{\LoveSignMean}{No}
\newcommand{\LoveMag}{0.8}
\newcommand{\LoveMagMean}{2.48}
\newcommand{\BetaWrittenLove}{2.5}

\newcommand{\BetaBidOne}{0.35}
\newcommand{\cdBidOne}{0.57}
\newcommand{\BidOnePvalString}{p<0.001}
\newcommand{\SEBidOne}{0.015}

\newcommand{\BidOneSig}{*}
\newcommand{\BidOneTestPval}{p<0.001}
\newcommand{\BidOneTestPvalMean}{p=0.064}
\newcommand{\BidOneSign}{Yes}
\newcommand{\BidOneSignMean}{Yes}
\newcommand{\BidOneMag}{1.43}
\newcommand{\BidOneMagMean}{3.65}
\newcommand{\BetaWrittenBidOne}{0.35}

\newcommand{\BetaBidTwo}{0.29}
\newcommand{\cdBidTwo}{0.47}
\newcommand{\BidTwoPvalString}{p<0.001}
\newcommand{\SEBidTwo}{0.015}

\newcommand{\BidTwoSig}{*}
\newcommand{\BidTwoTestPval}{p<0.001}
\newcommand{\BidTwoTestPvalMean}{p=0.053}
\newcommand{\BidTwoSign}{Yes}
\newcommand{\BidTwoSignMean}{Yes}
\newcommand{\BidTwoMag}{1.72}
\newcommand{\BidTwoMagMean}{4.36}
\newcommand{\BetaWrittenBidTwo}{0.29}

\newcommand{\BetaBidThree}{0.31}
\newcommand{\cdBidThree}{0.5}
\newcommand{\BidThreePvalString}{p<0.001}
\newcommand{\SEBidThree}{0.015}

\newcommand{\BidThreeSig}{*}
\newcommand{\BidThreeTestPval}{p<0.001}
\newcommand{\BidThreeTestPvalMean}{p=0.056}
\newcommand{\BidThreeSign}{Yes}
\newcommand{\BidThreeSignMean}{Yes}
\newcommand{\BidThreeMag}{1.61}
\newcommand{\BidThreeMagMean}{4.09}

\newcommand{\BetaConvict}{521.53}
\newcommand{\cdConvict}{0.16}
\newcommand{\ConvictPvalString}{p=0.012}
\newcommand{\SEConvict}{206.567}

\newcommand{\ConvictSig}{*}
\newcommand{\ConvictTestPval}{p<0.001}
\newcommand{\ConvictTestPvalMean}{p<0.001}
\newcommand{\ConvictSign}{Yes}
\newcommand{\ConvictSignMean}{Yes}
\newcommand{\ConvictMag}{9.59}
\newcommand{\ConvictMagMean}{3.42}
\newcommand{\BetaWrittenConvict}{521.53}

\newcommand{\BetaCases}{-74.632}

\newcommand{\SECases}{109.263}

\newcommand{\CasesSig}{\phantom{{*}}}
\newcommand{\CasesTestPval}{p=0.252}
\newcommand{\CasesTestPvalMean}{p<0.001}
\newcommand{\CasesSign}{Yes}
\newcommand{\CasesSignMean}{No}
\newcommand{\CasesMag}{2.68}
\newcommand{\CasesMagMean}{8.63}

\newcommand{\BetaRemorse}{-1153.061}
\newcommand{\cdRemorse}{-0.12}
\newcommand{\RemorsePvalString}{p=0.056}
\newcommand{\SERemorse}{603.325}

\newcommand{\RemorseSig}{\phantom{{*}}}
\newcommand{\RemorseTestPval}{p=0.002}
\newcommand{\RemorseTestPvalMean}{p=0.09}
\newcommand{\RemorseSign}{Yes}
\newcommand{\RemorseSignMean}{Yes}
\newcommand{\RemorseMag}{2.6}
\newcommand{\RemorseMagMean}{0.76}

\newcommand{\cdCasesRemorse}{-0.32}

\newcommand{\CasesRemorseIntPvalString}{p=0.047}

\newcommand{\BetaBar}{0.75}
\newcommand{\cdBar}{0.78}
\newcommand{\BarPvalString}{p<0.001}
\newcommand{\SEBar}{0.068}

\newcommand{\BarSig}{*}
\newcommand{\BarTestPval}{p=0.03}
\newcommand{\BarTestPvalMean}{p=0.998}
\newcommand{\BarSign}{Yes}
\newcommand{\BarSignMean}{Yes}
\newcommand{\BarMag}{0.8}
\newcommand{\BarMagMean}{0.54}

\newcommand{\BetaHeight}{0.003}

\newcommand{\SEHeight}{0.003}

\newcommand{\HeightSig}{\phantom{{*}}}
\newcommand{\HeightTestPval}{p<0.001}
\newcommand{\HeightTestPvalMean}{p=0.999}
\newcommand{\HeightSign}{Yes}
\newcommand{\HeightSignMean}{Yes}
\newcommand{\HeightMag}{33.33}
\newcommand{\HeightMagMean}{36}

\newcommand{\BetaFriend}{-0.002}

\newcommand{\SEFriend}{0.005}

\newcommand{\FriendSig}{\phantom{{*}}}
\newcommand{\FriendTestPval}{p<0.001}
\newcommand{\FriendTestPvalMean}{p=0.999}
\newcommand{\FriendSign}{No}
\newcommand{\FriendSignMean}{No}
\newcommand{\FriendMag}{100}
\newcommand{\FriendMagMean}{118}

\title{\vspace*{-1cm} \Large Automated Social Science: \\ Language Models as Scientist and Subjects\thanks{
  \footnotesize
  Thanks to generous support from Drew Houston and his AI for Augmentation and Productivity seed grant.
  Thanks to Jordan Ellenberg, Benjamin Lira Luttges, David Holtz, Bruce Sacerdote, Paul Röttger, Mohammed Alsobay, Ray Duch, Matt Schwartz, David Autor, and Dean Eckles for their helpful feedback.
  Author's contact information, code, and data are currently or will be available at \url{http://www.benjaminmanning.io/}.
}
}
\author{ \normalsize Benjamin S. Manning\thanks{Both authors contributed equally to this work.}  \\ \normalsize MIT \and
\normalsize Kehang Zhu\footnotemark[2] \\  \normalsize Harvard \and 
\normalsize John J. Horton   \\ \normalsize MIT \& NBER }
\date{\normalsize\today} 

\begin{document}

\maketitle

\begin{abstract}
\noindent We present an approach for automatically generating and testing, \emph{in silico}, social scientific hypotheses.
This automation is made possible by recent advances in large language models (LLM), but the key feature of the approach is the use of structural causal models. 
Structural causal models provide a language to state hypotheses, a blueprint for constructing LLM-based agents, an experimental design, and a plan for data analysis.
The fitted structural causal model becomes an object available for prediction or the planning of follow-on experiments.
We demonstrate the approach with several scenarios: a negotiation, a bail hearing, a job interview, and an auction.
In each case, causal relationships are both proposed and tested by the system, finding evidence for some and not others.
We provide evidence that the insights from these simulations of social interactions are not available to the LLM purely through direct elicitation.
When given its proposed structural causal model for each scenario, the LLM is good at predicting the signs of estimated effects, but it cannot reliably predict the magnitudes of those estimates.
In the auction experiment, the \emph{in silico} simulation results closely match the predictions of auction theory, but elicited predictions of the clearing prices from the LLM are inaccurate.
However, the LLM's predictions are dramatically improved if the model can condition on the fitted structural causal model.
In short, the LLM knows more than it can (immediately) tell. 
\end{abstract}

\newpage

\onehalfspacing

\section{Introduction} \label{sec:introduction}

There is much work on efficiently estimating econometric models of human behavior but comparatively little work on efficiently generating and testing those models to estimate.
Previously, developing such models and hypotheses to test was exclusively a human task. 
This is changing as researchers have begun to explore automated hypothesis generation through the use of machine learning.\footnote{
  A few examples include generative adversarial networks to formulate new hypotheses \citep{MLhypothesis2023}, algorithms to find anomalies in formal theories \citep{anomolies2023}, reinforcement learning to propose tax policies \citep{zheng2022ai}, random forests to identify heterogenous treatment effects \citep{CausalForests2018Athey}, and several others \citep{peterson2021lottery, EnkeComplexity2023, Buyalskaya2023Habits, Design2023Malone, girotra2023ideas}.
  }
But even with novel machine-generated hypotheses, there is still the problem of testing.
A potential solution is simulation.
Researchers have shown that Large Language Models (LLM) can simulate humans as experimental subjects with surprising degrees of realism.\footnote{\citep{park2023generative, bubeck2023sparks, argyle2023out, aher2023using, cogpsychgpt2022, brand2023using, AgreeBakker2022, fish2023generative, Jackson2024Turing}}
To the extent that these simulation results carry over to human subjects in out-of-sample tasks, they provide another option for testing \citep{horton2023large}.
In this paper, we combine these ideas---automated hypothesis generation and automated \emph{in silico} hypothesis testing---by using LLMs for both purposes.
We demonstrate that such automation is possible.
We evaluate the approach by comparing results to a setting where the real-world predictions are well known and test to see if an LLM can be used to generate information that it cannot access through direct elicitation.

The key innovation in our approach is the use of structural causal models to organize the research process.
Structural causal models are mathematical representations of cause and effect \citep{pearl2009causality, wright1934method} and have long offered a language for expressing hypotheses.\footnote{
  In an unfortunate clash of naming conventions, some disciplines have alternative definitions for the term ``structural'' when discussing formal models.
  Here, structural does not refer to the definition traditionally used in economics.
  See Appendix \ref{sec:scms} for a more detailed explanation.
}
What is novel in our paper is the use of these models as a blueprint for the design of agents and experiments.
In short, each explanatory variable describes something about a person or scenario that has to vary for the effect to be identified, so the system ``knows'' it needs to generate agents or scenarios that vary on that dimension---a straightforward transition from stated theory to experimental design and data generation.
Furthermore, the structural causal model offers a pre-specified plan for estimation \citep{haavelmo1944probability, haavelmo1943statistical,joreskogSEM1970}.

We built an open-source computational system implementing this structural causal model-based approach.
The system can automatically generate hypotheses, design experiments, run those experiments on independent LLM-powered agents, and analyze the results.
We use this system to explore several social scenarios: (1) two people bargaining over a mug, (2) a bail hearing for tax fraud, (3) a lawyer interviewing for a job, and (4) an open ascending price auction with private values for a piece of art.
We allow the system to propose the hypotheses for the first two scenarios and then run the experimental simulations without intervention.
For (3) and (4), we demonstrate the system's ability to accommodate human input at any point by selecting the hypotheses ourselves and editing some of the agents, but otherwise, we allow the system to proceed autonomously.

Though yet to be optimized for novelty, the system formulates and tests multiple falsifiable hypotheses---from which it generates several findings.
The probability of a deal increased as the seller's sentimental attachment to the mug decreased, and both the buyer's and the seller's reservation prices mattered.
A remorseful defendant was granted lower bail but was not so fortunate if his criminal history was extensive.
However, the judge's case count before the hearing---which was hypothesized to matter---did not affect the final bail amount.
The candidate passing the bar exam was the only important factor in her getting the job.
Neither the candidate's height nor the interviewer's friendliness affected the outcome.

The auction scenario is particularly illuminating.
An increase in the bidders' reservation prices caused an increase in the clearing price, a clearing price that is always close to the second-highest reservation amongst the bidders.
These simulation results closely match the theory \citep{PrivateAuctionMaskin1985} and what has been observed empirically \citep{Athey2011Auction}.

None of the findings from the system's experiments are ``counterintuitive,'' but it is important to emphasize they were the result of empiricism, not just model introspection.
However, this does raise the question of whether the simulations are even necessary.\footnote{
  Performing these experiments required a substantial software infrastructure.
}
Instead of simulation, could an LLM simply do a ``thought experiment'' about the proposed \emph{in silico} experiment and achieve the same insight?
To test this idea, we describe the experiments that will be simulated and ask the LLM to predict the results---both the path estimates and point predictions.
The path estimates being the coefficients in the linear structural causal model.
To make this concrete, suppose we had the simple linear model $y = X \beta $ to describe some scenario, and we ran an experiment to estimate $\hat{\beta}$.
We describe the scenario and the experiment to the LLM and ask it to predict $y_i$ given a particular $X_i$ (a ``predict-$y_i$'' task).
Separately, we ask it to predict $\hat{\beta}$ (a ``predict-$\hat{\beta}$'' task).
Later, we examine how the LLM does on the predict-$y_i$ task when it has access to the fitted structural causal model (i.e., $\hat{\beta}$).

In the predict-$y_i$ task, we prompt the LLM to predict the outcome $y_i$ given each possible combination of the $X_i$'s from the auction experiment.
Direct elicitation of the predictions for $y_i$ in the auction experiment is wildly inaccurate.
The predictions are even further from the theory than the empirical results.

In the predict-$\hat{\beta}$ task, the LLM is asked to predict the fitted structural causal model's path estimates for all four experiments, provided with contextual information about each scenario. 
On average, the LLM predicts the path estimates are \MeanMagPred \space times larger than the experimental results. 
Its predictions are overestimates for 10 out of 12 of the paths, although they are generally in the correct direction.

We repeat the predict-$y_i$ task, but this time, we provide the LLM with the experimental path estimates.
For each $X_i$, we fit the structural causal model using all but the $i$th observation and then ask the LLM to predict $y_i$ given $X_i$ and this fitted model.
In this ``predict-$y_i|\hat{\beta}_{-i}$'' task, the predictions are far better than in the predict-$y_i$ task without the fitted model.
The mean squared error is six times lower, and the predictions are much closer to those made by the theory, but they are still further from the theory than they are to the simulations.

We design and implement an approach to automated social science because LLMs possess latent information about human behavior that can be systematically explored and extracted \citep{burns2023discovering, scherrer2024evaluating}.
These models are trained to predict the next token in a sequence of text from a massive human-generated corpus.
From this straightforward objective, the models develop a remarkably sophisticated model of the world, at least as captured in text \citep{gurnee2023timespace, bubeck2023sparks, patel2021mapping}.
And while there are many situations where LLMs are imperfect proxies for humans \citep{santurkar2023opinions,Cheng2023caricature}, there is also a growing body of work demonstrating that experiments with LLMs as subjects can predict human behavior in never-before-seen tasks \citep{ValidityLLM2024, binz2023turning}. 
Rapid and automated exploration of these models' behavior could be a powerful tool to efficiently generate new insights about humans.
Our contribution is to demonstrate that it is possible to create such a tool: a system that can simulate the entire social scientific process without human input at any step.

The remainder of this paper is structured as follows: 
Section \ref{sec:overview} provides an overview of the system.
Section \ref{sec:results} provides some results generated using our system.
Section \ref{sec:predictions} explores an LLM's capacity to predict the results in Section \ref{sec:results}.
Section \ref{sec:advantages} discusses the advantages of using SCMs over other methods for studying causal relationships in simulations of social interactions.
The paper concludes in Section \ref{sec:conclusion}.

\section{Overview of the system} \label{sec:overview}

To perform this automated social science, we needed to build a system.
The system intentionally mirrors the experimental social scientific process.
These steps are, in broad strokes: 

\begin{enumerate}[itemsep=0pt]
\item	Social scientists start by selecting a topic or domain to study (e.g., misinformation, auctions, bargaining, etc).
\item Within the domain, they identify interesting outcomes and some causes that might affect the outcomes.
These variables and their proposed relationships are the hypotheses. 
\item They design an experiment to test these hypotheses by inducing variation in the causes and measuring the outcomes. 
\item After designing the experiment, social scientists determine how they will analyze the data in a pre-analysis plan.
\item Next, they recruit participants, run the experiment, and collect the data.
\item Finally, they analyze the data per the pre-analysis plan to estimate the relationships between the proposed causes and outcomes.
\end{enumerate}
While any given social scientist might not follow this sequence exactly, whatever their approach may be, the first two steps should always guide the later steps---the development of the hypothesis guides the experimental design and model estimation.
Of course, many social scientists must often omit steps 3-5 when a controlled experiment is not possible, but they typically have some notion of the experiment they would like to run.

To build our system, we formalized a sequence of these steps analogous to those listed above.
The system executes them autonomously.
Since the system uses AI agents instead of human subjects, it can \emph{always} design and execute an experiment.

Structural causal models (SCM) are essential to the design of the system because they make unambiguous causal statements, which allow for unambiguous estimation and experimental design.\footnote{
  We use simple linear SCMs unless stated otherwise.
  This assumption is not necessarily correct but offers an unequivocal starting point to generate hypotheses.
  Functional assumptions can be tested by comparing fitted SCMs with various forms using data generated from a known causal structure.
  Section \ref{sec:scms} in the appendix provides a more detailed explanation of SCMs.
}
Algorithms can determine precisely which variables must be exogenously manipulated to identify the effect of a given cause \citep{pearl2009causality}.
If the first two steps in the social scientific process are building the SCM, the last four can be directly determined subject to the SCM.
Such precision makes automation possible as the system only relies on a few key early decisions.
Otherwise, the space of possible choices for the latter steps would explode, making automation infeasible.

The system is implemented in Python and uses GPT-4 for all LLM queries.
Its decisions are editable at every step.
The overview in this section is a high-level description of the system, but there are many more specific design choices and programming details in Appendix \ref{sec:implementation}.
For the purposes of most readers, the high-level overview should be sufficient to understand the system's process, the results we present in Section \ref{sec:results}, and the additional analyses in Sections \ref{sec:predictions} and \ref{sec:advantages}.

\begin{figure}[ht]
  \caption{An overview of the automated system.}
     \label{fig:system_overview}
    \centering
    \includegraphics[width=\linewidth]{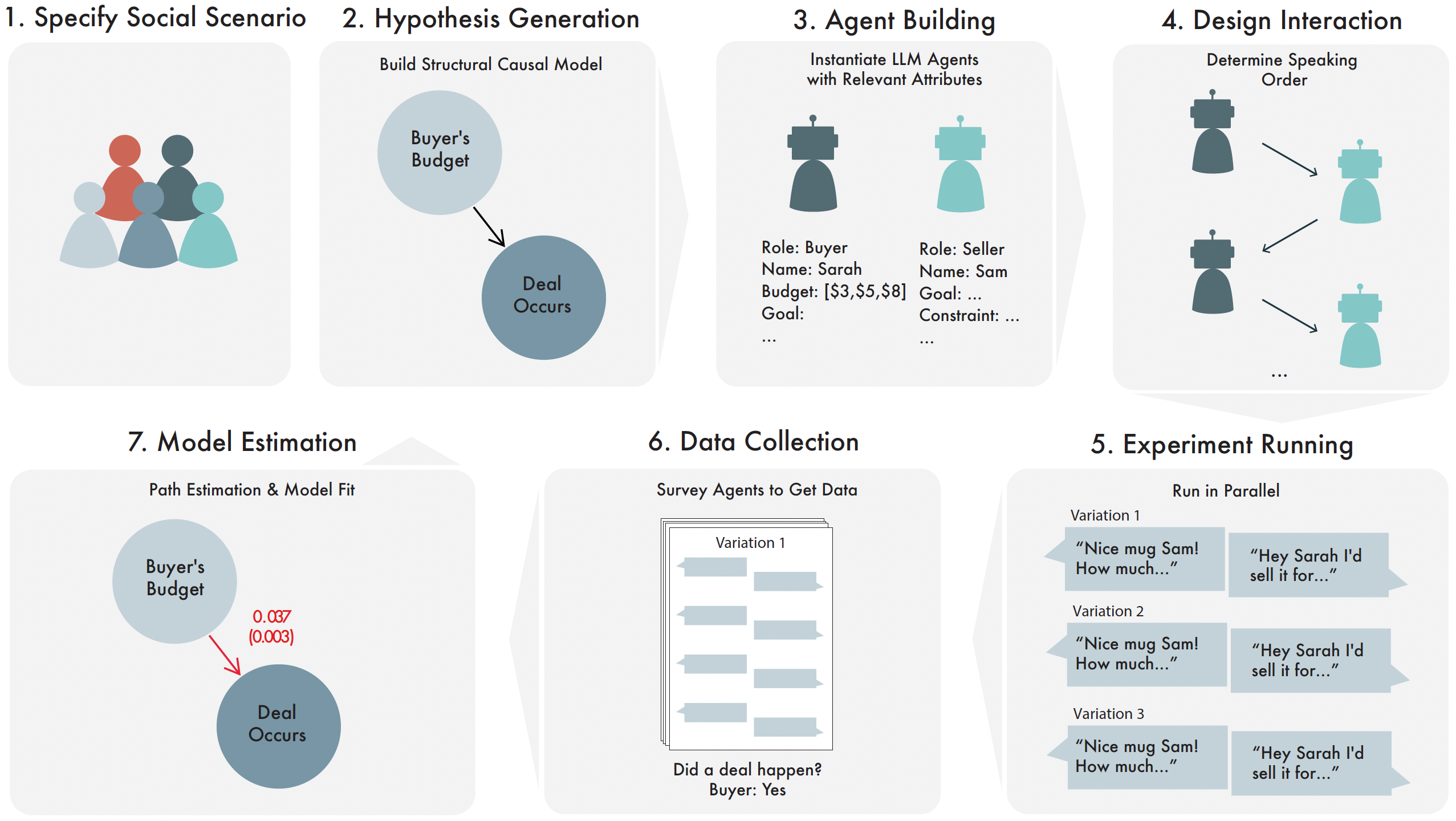}
    \begin{minipage}{\textwidth}
    \begin{footnotesize}
  \emph{Notes: 
  Each step in the process corresponds to an analogous step in the social scientific process as done by humans. 
  The development of the hypothesis guides the experimental design, execution, and model estimation.
  Researchers can edit the system's decisions at any step in the process.}
    \end{footnotesize}
    \end{minipage}
\end{figure}

The system takes as input some scenario of social scientific interest: a negotiation, a bail decision, a job interview, an auction, and so on. 
Starting with (1) this input, the system (2) generates outcomes of interest and their potential causes, (3) creates agents that vary on the exogenous dimensions of said causes, (4) designs an experiment, (5) executes the experiment with LLM-powered agents simulating humans, (6) surveys the agents to measure the outcomes, (7) analyzes the results of the experiment to assess the hypotheses, which can be used to plan a follow-on experiment.
Figure~\ref{fig:system_overview} illustrates these steps, and we will briefly explore each in greater depth.

The first step is to generate hypotheses as SCMs based on the social scenario, the scenario being the only necessary input to the system.
This is done by querying an LLM for the relevant agents and then interesting outcomes, their potential causes, and methods to operationalize and measure both.\footnote{
  When we say ``query an LLM,'' we mean this literally.
  We have written a prompt that the system provides to an LLM with the scenario.
  For example, the prompt used to generate the relevant agents is:
  \emph{In the following scenario: ``\{\emph{scenario}\}'', who are the individual human agents in a simple simulation of this scenario?}
  Where ``\{scenario\}'' is replaced with the scenario of interest.
  The LLM then returns a list of agents, which are stored in the system and can be used in follow-on prompts, prompts that generate things like the outcomes and proposed causes.
  The system contains over 50 pre-written scenario-neutral prompts to gather all the information needed to generate the SCM, run the experiment, and analyze the results.
}
We use \texttt{Typewriter text} to indicate example output from the system.
Suppose the social scenario is ``two people bargaining over a mug.'' 
The LLM may generate \texttt{whether a deal occurs for the mug} as an outcome, and operationalizes the outcome as \texttt{a binary variable with a ``1'' when a deal occurs and a ``0'' when it does not}.
It then generates potential exogenous causes and their operationalizations: the \texttt{buyer's budget}, which is operationalized as \texttt{the buyer's willingness to pay in dollars}.
The system takes each of these variables, constructs an SCM (see the second step in Figure~\ref{fig:system_overview}), and stores the relevant information about the operationalizations associated with each variable.\footnote{
  The system generates several other pieces of information about each variable, which help guide the experimental design and data analysis.
  See Appendix \ref{sec:implementation} for further details.}\footnote{
    The graph in the second step of Figure~\ref{fig:system_overview} is a directed acyclic graph (DAG).
    For convenience, we will use DAGs to represent SCMs throughout the paper and assume they imply a simple linear model unless stated otherwise.
  }
From this point on, the SCM serves as a blueprint for the rest of the process, namely the automatic instantiation of agents, their interaction, and the estimation of the linear paths.

The second step is to construct the relevant agents---the \texttt{Buyer} and the \texttt{Seller} in Figure~\ref{fig:system_overview}, step 3.
By ``construct,'' we mean that the system prompts independent LLMs to be people with sets of attributes.
These attributes are the exogenous dimensions of the SCM, dimensions that are varied in each simulation.
I.e., the different experimental conditions.
For the current scenario, a \texttt{Budget} is provided to the buyer that can take on values of \texttt{\{\$5, \$10, \$20, \$40\}}.
By simulating interactions of agents that vary on the exogenous dimensions of the SCM, the data generated can be used to fit the SCM.

Next, the system generates survey questions to gather data about the outcomes from the agents automatically once each simulation is complete.
An LLM can easily generate these questions when provided with information about the variables in the SCM (e.g., asking the buyer, ``\texttt{Did a deal happen?}'').
All LLM-powered agents in our system have ``memory.'' 
They store what happened during the simulation in text, making it easy to ask them questions about what happened.

Fourth, the system determines how the agents should interact.
LLMs are designed to generate text in sequence.
Since independent LLMs power each agent, one agent must finish speaking before the next begins.
This necessitates a turn-taking protocol to simulate the conversation.
We programmed a menu of six ordering protocols, from which an LLM is queried to select the most appropriate for a given scenario.
We describe each protocol in Appendix~\ref{sec:implementation}, and they are presented in Figure~\ref{fig:interaction_types}, but in our bargaining scenario with two agents, there are only two possible ways for the agents to alternate speaking. 
In this case, the system selects: \texttt{speaking order: (1) Buyer, (2) Seller}, (step 4, Figure~\ref{fig:system_overview}).
The speaking order can be flexible in more complex simulations with more agents, such as an auction or a bail hearing.

Now, the system runs the experiment.
The conditions are simulated in parallel (step 5 in Figure~\ref{fig:system_overview}), each with a different value for the exogenous dimensions of the SCM---the possible budgets for the buyer.

The system must also determine when to stop the simulations.
There is no obvious rule for when a conversation should end.
Like the halting problem in computer science---it is impossible to write a universal algorithm that can determine whether a given program will complete \citep{turing1936Halting}---such a rule for conversations does not exist.
We set two stopping conditions for the simulations.
After each agent speaks in a simulation, an external LLM is prompted with the transcript of the conversation and asked if the conversation should continue.
If yes, the next agent speaks; otherwise, the simulation ends.
Additionally, we limit the total number of agent statements to twenty.
One could imagine doing something more sophisticated both with the social interactions and the stopping conditions in the future.
This is even a place for possible experimentation as the structure of social interactions can impact various outcomes of interest \citep{WeakTiesAral2022,SacerdotePeer2001, LongTies2023Jahani}.

Finally, the system gathers the data for analysis.
Outcomes are measured by asking the agents the survey questions (Figure~\ref{fig:system_overview}, step 6) as determined before the experiment.
The data is then used to estimate the linear SCM.
For our negotiation, that would be a simple linear model with a single path estimate (i.e., linear coefficient) for the effect of the buyer's budget on the probability of a deal---the final step in Figure~\ref{fig:system_overview}.
Note that an SCM specifies, ex-ante, the exact statistical analyses to be conducted after the experiment---akin to a pre-analysis plan.
This step of the system's process is, therefore, mechanical.

The system, as outlined, is automated from start to finish---the SCM and its accompanying metadata serve as a blueprint for the rest of the process.
Once there is a fitted SCM, this process can be repeated.
Although we have not automated the transition from one experiment to the next, the system can generate new causal variables, induce variations, and run another experiment based on the results of the first.

\section{Results of experiments}  \label{sec:results}

We present results for four social scenarios explored using the system.
In the first two scenarios, our involvement in the system's process was restricted to entering the description of the scenario and then the entire process was automated.
In the third and fourth scenarios, we selected the hypotheses and edited some of the agents, but the system designed and executed the experiments.
We intervened in the latter scenarios not because the system is incapable of simulating these scenarios autonomously, but to demonstrate the system's capacity to accommodate human input at any point while still generating exciting results.

\subsection{Bargaining over a mug} \label{sec:mug-results}

We first use the system to simulate ``two people bargaining over a mug''---this phrase being in quotes because it was the only input needed for the system to simulate the following process.
The system selected a buyer and seller as the relevant agents, the outcome as whether a deal occurs, and the buyer's budget, the seller's minimum acceptable price, and the seller's emotional attachment to the mug as potential causes.

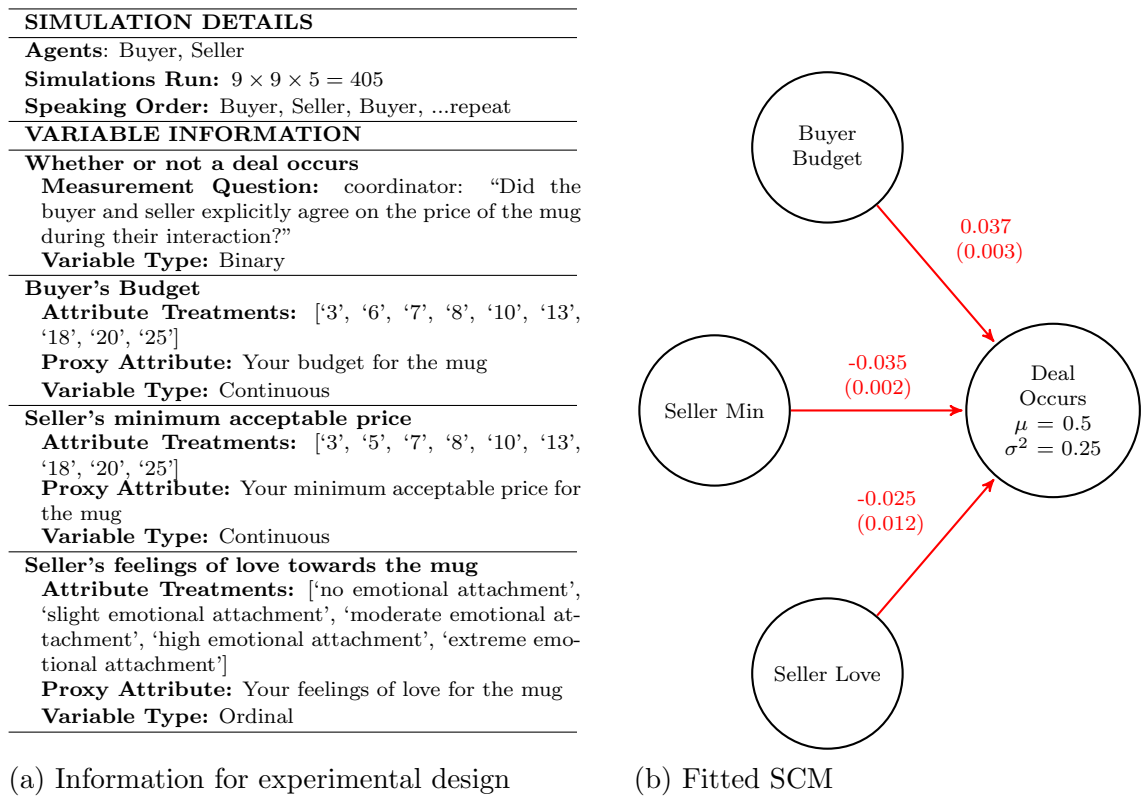
\begin{figure}[htbp]
  \caption{Experimental design and fitted SCM for “two people bargaining over a mug.”}
  \label{fig:mug-love}
  \begin{subfigure}[b]{.5\textwidth}
    \scriptsize
    \renewcommand{\arraystretch}{1.1}
    \begin{tabularx}{\textwidth}{lX}
      \hline
      \textbf{SIMULATION DETAILS} \\
      \hline
      \textbf{Agents}: Buyer, Seller \\ 
      \parbox{.95\textwidth}{\textbf{Simulations Run: $9 \times 9 \times 5 = 405$}} \\
      \textbf{Speaking Order:} Buyer, Seller, Buyer, ...repeat\\
      \hline
      \textbf{VARIABLE INFORMATION}  \\
      \hline
      \textbf{Whether or not a deal occurs} \\ \hspace{.75em}\parbox{.95\textwidth}{\textbf{Measurement Question:} coordinator: ``Did the buyer and seller explicitly agree on the price of the mug during their interaction?''} \\ \hspace{.75em}\parbox{.95\textwidth}{\textbf{Variable Type:} Binary} \\
      \hline
      \textbf{Buyer's Budget} \\ \hspace{.75em}\parbox{.95\textwidth}{\textbf{Attribute Treatments:} [`3', `6', `7', `8', `10', `13', `18', `20', `25']} \\ 
      \hspace{.75em}\parbox{.95\textwidth}{\textbf{Proxy Attribute:} Your budget for the mug} \\ 
      \hspace{.75em}\parbox{.95\textwidth}{\textbf{Variable Type:} Continuous} \\
      \hline
      \textbf{Seller's minimum acceptable price} \\ \hspace{.75em}\parbox{.95\textwidth}{\textbf{Attribute Treatments:} [`3', `5', `7', `8', `10', `13', `18', `20', `25']} \\
      \hspace{.75em}\parbox{.95\textwidth}{\textbf{Proxy Attribute:} Your minimum acceptable price for the mug} \\
      \hspace{.75em}\parbox{.95\textwidth}{\textbf{Variable Type:} Continuous} \\
      \hline
      \textbf{Seller's feelings of love towards the mug} \\ \hspace{.75em}\parbox{.95\textwidth}{\textbf{Attribute Treatments:} [`no emotional attachment', `slight emotional attachment', `moderate emotional attachment', `high emotional attachment', `extreme emotional attachment']} \\
      \hspace{.75em}\parbox{.95\textwidth}{\textbf{Proxy Attribute:} Your feelings of love for the mug} \\
      \hspace{.75em}\parbox{.95\textwidth}{\textbf{Variable Type:} Ordinal} \\
      \hline \\
    \end{tabularx}
    \caption{Information for experimental design}
    \label{tab:mug-love-table}
\end{subfigure}
\begin{subfigure}[b]{.45\textwidth}
  \centering
  \begin{tikzpicture}[->,>=stealth',shorten >=1pt,auto,node distance=3cm,
                      thick,
                      main node/.style={circle, draw, font=\scriptsize, text width=1.5cm, align=center, minimum size=2cm}]
    \node[main node] (A) at (0,0) {Deal Occurs \\ $\mu =\MeanDeal$ \\ $\sigma^2 = \VarDeal$};
    \node[main node] (B) at (-3,3.5) {Buyer Budget};
    \node[main node] (C) at (-4.5,0) {Seller Min};
    \node[main node] (D) at (-3,-3.5) {Seller Love};
   \path[every node/.style={font=\scriptsize}]
      (B) edge[red] node[align=center, text=red] {\BetaBudget\\ \phantom{-}(\SEBudget)} (A)
      (C) edge[red] node[align=center, text=red] {\BetaMinPrice\\(\SEMinPrice)}(A)
      (D) edge[red] node[align=center, text=red] {\BetaLove \\\phantom{-}(\SELove)} (A);
  \end{tikzpicture}
  \caption{Fitted SCM}
  \label{fig:mug-love-scm}
\end{subfigure}
  \begin{minipage}{\textwidth}
    \begin{footnotesize}
    \emph{Notes: Figure~\ref{tab:mug-love-table} provides the information automatically generated by the system to execute the experiment for its proposed hypothesis.
    This includes the high level structure of the simulations, how the outcome is measured, and the treatment variations for each of the causes.
    The fitted SCM in Figure~\ref{fig:mug-love-scm} shows the results of the experiment.
    The outcome is given with its mean and variance.
    The edges are labeled with their unstandardized path estimate and standard error.
    We assume a simple linear model for the SCM, such that the above graph can also be written as $DealOccurs = \BetaBudget BuyerBudget \BetaMinPrice MinPrice  \BetaLove SellerLove$.
    }
    \end{footnotesize}
    \end{minipage}
\end{figure}

Table~\ref{tab:mug-love-table} provides the information generated by the system about the SCM and the experimental design.
The topmost row, simulation details, provides high-level information about the structure of the simulation.
The remaining rows provide information about the variables in the SCM and how they were operationalized.
The system automatically generated all this information by iteratively querying the LLM.

The three exogenous variables were operationalized as the buyer's budget in dollars, the seller's minimum acceptable price in dollars, and the seller's emotional attachment as an ordinal scale from ``no emotional attachment'' to ``extreme emotional attachment.''
The system chose nine values (the ``Attribute Treatments'' in Table~\ref{tab:mug-love-table}) to vary for each of the first two causes and five for the seller's feelings of love towards the mug (one for each level of the scale).
This led to $9 \times 9 \times 5 = 405$ experimental runs of the simulated conversation between the buyer and seller.

Figure \ref{fig:mug-love-scm} provides the fitted SCM.
The outcome variable is given with its mean and variance.
The raw path estimates and their standard errors are shown on the arrows.
For ordinal variables (e.g., the seller's feelings of love), we treat the levels as numerical values.
The buyer and seller reached a deal for the mug in half of the simulations, and all three causes had a statistically significant effect on the probability of a deal.

A one-dollar increase in the buyer's budget caused an average increase of \BetaWrittenBudget \space percentage points in the probability of a deal ($\hat{\beta}\text{*} = \cdBudget$, $\BudgetPvalString$).\footnote{
  We report standardized effect size estimates with $\hat{\beta}$*. 
  Standardized effect sizes being ``a one standard deviation increase in X causes a $\hat{\beta}$* standard deviation increase in Y.''
}
A one-dollar increase in the seller's minimum acceptable price caused an average decrease of \BetaWrittenMinPrice \space percentage points in the probability of a deal occurring ($\hat{\beta}\text{*} = \cdMinPrice$, $\MinPricePvalString$).
Finally, a one-unit increase in the ordinal scale of the seller's love for the mug, such as going from moderate emotional attachment to high emotional attachment, caused an average decrease of \BetaWrittenLove \space percentage points in the probability of a deal ($\hat{\beta}\text{*} = \cdLove$, $\LovePvalString$).

\subsection{A bail hearing}

\begin{figure}[htbp]
  \caption{Experimental design and fitted SCM for “a judge is setting bail for a criminal defendant who committed 50,000 dollars in tax fraud.”}
  \label{fig:mug-love}
  \begin{subfigure}[b]{.5\textwidth}
    \scriptsize
    \renewcommand{\arraystretch}{1.1}
    \begin{tabularx}{\textwidth}{lX}
      \hline
      \textbf{SIMULATION DETAILS} \\
      \hline 
      \textbf{Agents}: Judge, Defendant, Defense attorney, Prosecutor \\ 
      \parbox{.95\textwidth}{\textbf{Simulations Run: $7 \times 7 \times 5 = 243$}} \\
      \textbf{Speaking Order:} Judge, Prosecutor, Judge,\\\hspace{.75em} Defense Attorney, Judge, Defendant, ... repeat\\
      \hline
      \textbf{VARIABLE INFORMATION}  \\
      \hline
      \textbf{Bail amount set by the judge} \\ \hspace{.75em}\parbox{.95\textwidth}{\textbf{Measurement Question:} Judge: ``What was the bail amount you set for the defendant?''} \\ \hspace{.75em}\parbox{.95\textwidth}{\textbf{Variable Type:} Continuous} \\
      \hline
      \textbf{Defendant's criminal history} \\ \hspace{.75em}\parbox{.95\textwidth}{\textbf{Attribute Treatments:} [`0', `1', `2', `3', `6', `9', `12']} \\ 
      \hspace{.75em}\parbox{.95\textwidth}{\textbf{Proxy Attribute:} Number of your prior convictions} \\ 
      \hspace{.75em}\parbox{.95\textwidth}{\textbf{Variable Type:} Count} \\
      \hline
      \textbf{Prior case count for judge that day} \\ \hspace{.75em}\parbox{.95\textwidth}{\textbf{Attribute Treatments:} [`0', `2', `5', `9', `12', `18', `23']} \\
      \hspace{.75em}\parbox{.95\textwidth}{\textbf{Proxy Attribute:} Number of cases you have already heard today} \\
      \hspace{.75em}\parbox{.95\textwidth}{\textbf{Variable Type:} Count} \\
      \hline
      \textbf{Defendant's level of remorse} \\ \hspace{.75em}\parbox{.95\textwidth}{\textbf{Attribute Treatments:} [`no expressed remorse', `low expressed remorse', `moderate expressed remorse', `high expressed remorse', `extreme expressed remorse']} \\
      \hspace{.75em}\parbox{.95\textwidth}{\textbf{Proxy Attribute:} Your level of expressed remorse} \\
      \hspace{.75em}\parbox{.95\textwidth}{\textbf{Variable Type:} Ordinal} \\
      \hline \\
    \end{tabularx}
    \caption{Information for experimental design}
    \label{tab:tax-fraud-table}
\end{subfigure}
\begin{subfigure}[b]{.45\textwidth}
  \centering
  \begin{tikzpicture}[->,>=stealth',shorten >=1pt,auto,node distance=3cm,
                      thick,
                      main node/.style={circle, draw, font=\scriptsize, text width=1.5cm, align=center, minimum size=2cm}]
    % nodes
    \node[main node] (A) at (0,0) {Bail Amount \\ $\mu =\MeanBail$ \\ $\sigma^2 = 1.9e7$};
    \node[main node] (B) at (-3,3.5) {Criminal History};
    \node[main node] (C) at (-4.5,0) {Judge Case Count};
    \node[main node] (D) at (-3,-3.5) {Defendant's Remorse};
   % edges
   \path[every node/.style={font=\scriptsize}]
      (B) edge[red] node[align=center, text=red] {\BetaConvict\\ \phantom{-}(\SEConvict)} (A)
      (C) edge[red] node[align=center, text=red] {\BetaCases\\(\SECases)}(A)
      (D) edge[red] node[align=center, text=red] {\BetaRemorse \\\phantom{-}(\SERemorse)} (A);
  \end{tikzpicture}
  \caption{Fitted SCM}
  \label{fig:tax-fraud-scm}
\end{subfigure}
  \begin{minipage}{\textwidth}
    \begin{footnotesize}
    \emph{Notes: Figure~\ref{tab:tax-fraud-table} provides the information automatically generated by the system to execute the experiment for its proposed hypothesis.
    Figure~\ref{fig:tax-fraud-scm} shows the fitted SCM from the experiment.
    }
    \end{footnotesize}
    \end{minipage}
\end{figure}
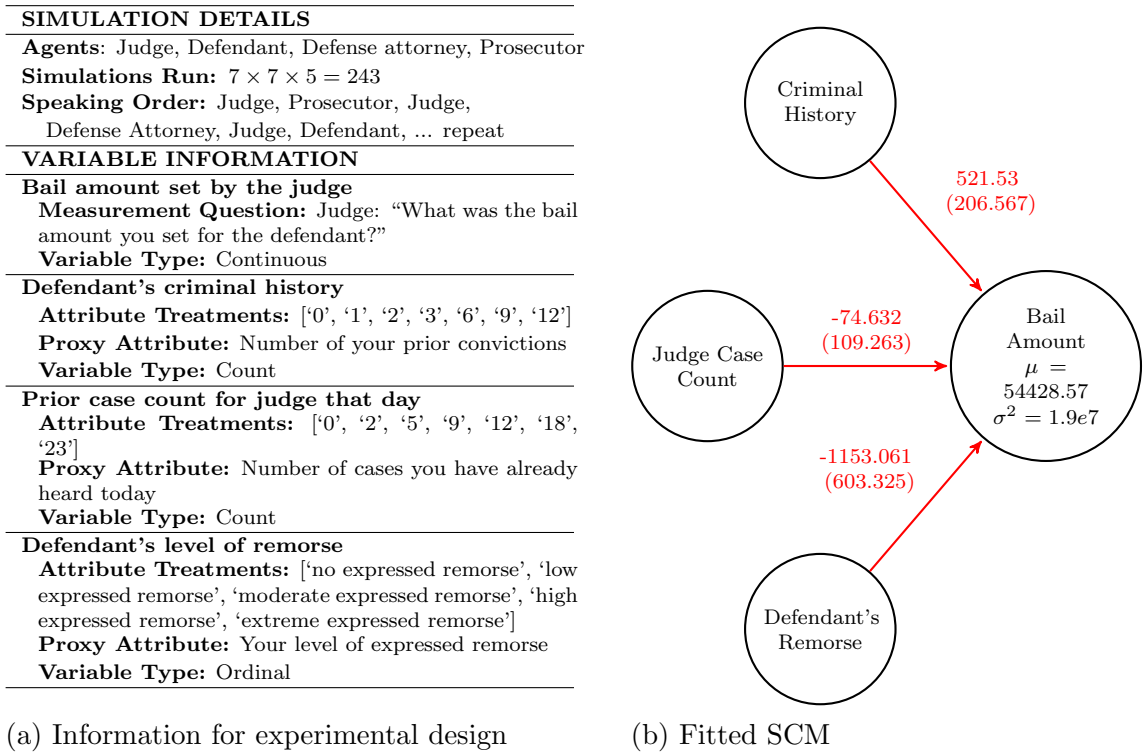

Next, we explore ``a judge is setting bail for a criminal defendant who committed 50,000 dollars in tax fraud.''
Table \ref{tab:tax-fraud-table} shows that the system selected a judge, defendant, defense attorney, and prosecutor as the relevant agents.
In this scenario, the system selected a more flexible interaction protocol than the one used in the previous experiment.
The judge was chosen as a center agent and, in order, the prosecutor, defense attorney, and defendant as the non-center agents.
This means the judge spoke first in every simulation, alternating with the other agents: judge, prosecutor, judge, defense attorney, judge, defendant, and so on.
As described in Section~\ref{sec:interactions}, we call this the ``center-ordered'' interaction protocol.

The system chose the outcome to be the final bail amount, and the three proposed causes are the defendant's criminal history, the number of cases the judge has already heard that day, and the defendant's level of remorse.
The number of cases the judge already heard that day and the defendant's level of remorse are operationalized literally, as the count of cases the judge has heard and five ordinal levels of possible outward expressions of remorsefulness.
The defendant's criminal history is operationalized as the number of previous convictions.

In the fitted SCM in Figure \ref{fig:tax-fraud-scm}, only the defendant's criminal history had a significant effect on the final bail amount with each additional conviction causing an average increase of \$\BetaWrittenConvict \space in bail ($\hat{\beta}\text{*} = \cdConvict$, $\ConvictPvalString$).
It is unclear whether the defendant's remorse affected the final bail amount.
The effect size was small but non-trivial with borderline significance ($\hat{\beta}\text{*}  = \cdRemorse$, and $\RemorsePvalString$).

When we estimated the SCM with interactions, the interaction between the judge's case count and the defendant's remorse was nontrivial ($\hat{\beta}\text{*}  = \cdCasesRemorse$, $\CasesRemorseIntPvalString$).
In this specification (Figure~\ref{fig:tax-fraud-interaction-scm}), none of the other interactions or the stand-alone causes have a significant effect, including the defendant's criminal history.

\subsection{Interviewing for a job as a lawyer}

In our third simulated experiment, we chose the scenario ``a person interviewing for a job as a lawyer.''
The system determined that a job applicant and an employer were the agents.
Unlike the previous simulations, we manually selected the variables in the SCM.
Table \ref{tab:lawyer-interview-3var-table} shows that these
were the employer's hiring decision as the outcome and whether the applicant passed the bar, the interviewer's friendliness, and the job applicant's height as the potential causes.

\begin{figure}[htbp]
  \caption{Experimental design and fitted SCM for “a person is interviewing for a job as a lawyer.”}
  \label{fig:lawyer_interview_3var}
  \begin{subfigure}[b]{.5\textwidth}
    \scriptsize
    \renewcommand{\arraystretch}{1.1}
    \begin{tabularx}{\textwidth}{lX}
      \hline
      \textbf{SIMULATION DETAILS} \\
      \hline
      \textbf{Agents}: Interviewer, Job Applicant \\ 
      \parbox{.95\textwidth}{\textbf{Simulations Run: $2 \times 5 \times 8 = 405$}} \\
      \textbf{Speaking Order:} Interviewer, Job Applicant,\\\hspace{.75em}Interviewer, ...repeat\\
      \hline
      \textbf{VARIABLE INFORMATION}  \\
      \hline
      \textbf{Employer's Decision} \\ \hspace{.75em}\parbox{.95\textwidth}{\textbf{Measurement Question:} Employer: ``Have you decided to hire the job applicant?''} \\ \hspace{.75em}\parbox{.95\textwidth}{\textbf{Variable Type:} Binary} \\
      \hline
      \textbf{Whether Applicant Passed Exam} \\ \hspace{.75em}\parbox{.95\textwidth}{\textbf{Attribute Treatments:} [`Passed', `Not']} \\ 
      \hspace{.75em}\parbox{.95\textwidth}{\textbf{Proxy Attribute:} Your bar exam status} \\ 
      \hspace{.75em}\parbox{.95\textwidth}{\textbf{Variable Type:} Binary} \\
      \hline
      \textbf{Interviewer's level of friendliness} \\ \hspace{.75em}\parbox{.95\textwidth}{\textbf{Attribute Treatments:} [`2', `7', `12', `17', `22']} \\
      \hspace{.75em}\parbox{.95\textwidth}{\textbf{Proxy Attribute:} Number of positive phrases to use during interview} \\
      \hspace{.75em}\parbox{.95\textwidth}{\textbf{Variable Type:} Count} \\
      \hline
      \textbf{Job applicant's height} \\ \hspace{.75em}\parbox{.95\textwidth}{\textbf{Attribute Treatments:} [`160', `165', `170', `175', `180', `185', `190', `195']} \\
      \hspace{.75em}\parbox{.95\textwidth}{\textbf{Proxy Attribute:} Your height in centimeters} \\
      \hspace{.75em}\parbox{.95\textwidth}{\textbf{Variable Type:} Continous} \\
      \hline \\
    \end{tabularx}
    \caption{Information for experimental design}
    \label{tab:lawyer-interview-3var-table}
\end{subfigure}
\begin{subfigure}[b]{.45\textwidth}
  \centering
  \begin{tikzpicture}[->,>=stealth',shorten >=1pt,auto,node distance=3cm,
                      thick,
                      main node/.style={circle, draw, font=\scriptsize, text width=1.5cm, align=center, minimum size=2cm}]
    \node[main node] (A) at (0,0) {Employer Decision \\ $\mu =\MeanHire$ \\ $\sigma^2 = \VarHire$};
    \node[main node] (B) at (-3,3.5) {Passed Bar};
    \node[main node] (C) at (-4.5,0) {Interviewer Friendliness};
    \node[main node] (D) at (-3,-3.5) {Applicant Height};
   \path[every node/.style={font=\scriptsize}]
      (B) edge[red] node[align=center, text=red] {\BetaBar\\ \phantom{-}(\SEBar)} (A)
      (C) edge[red] node[align=center, text=red] {\BetaFriend\\(\SEFriend)}(A)
      (D) edge[red] node[align=center, text=red] {\BetaHeight\\\phantom{-}(\SEHeight)} (A);
  \end{tikzpicture}
  \caption{Fitted SCM}
  \label{fig:lawyer-interview-3var-scm}
\end{subfigure}
  \begin{minipage}{\textwidth}
    \begin{footnotesize}
    \emph{Notes: Figure~\ref{tab:lawyer-interview-3var-table} provides the information automatically generated by the system to execute the experiment for the proposed hypothesis.
    Figure~\ref{fig:lawyer-interview-3var-scm} shows the fitted SCM from the experiment.
    }
    \end{footnotesize}
    \end{minipage}
\end{figure}

The system operationalized the causes as a binary variable for passing the bar, the job applicant's height in centimeters, and the interviewer's friendliness as the proposed number of friendly phrases to use during the simulation.
Since one of the causes is a binary variable, the only potential cause in all our scenarios of this type, the sample size for the experimental simulations of this scenario is smaller ($n=80$).
By default, the system runs a factorial experimental design for all proposed values of each cause.
With only two possible values for the job applicant passing the bar (as opposed to 5 varied treatment values for the interviewer's friendliness and 8 for the applicant's height), this limits the possible combinations of the causal variables to $2 \times 5 \times 8 = 80$.
A researcher could run more simulations to increase the sample size if so desired.

We can see in Figure \ref{fig:lawyer-interview-3var-scm} that only the applicant passing the bar has a clear causal effect on whether the applicant gets the job. 
This is the largest standardized effect we see across the simulations in the four scenarios ($\hat{\beta}\text{*}  = \cdBar$, $\BarPvalString$).
On average, whether or not the applicant passes the bar increases the probability she gets the job by 75 percentage points. 
When we test for interactions, none are significant (Figure \ref{fig:lawyer-interview-3var-interaction-scm}).

\subsection{An auction for a piece of art}

\begin{figure}[htbp]
  \caption{Experimental design and fitted SCM for “3 bidders participating in an auction for a piece of art starting at fifty dollars.”}
  \label{fig:auction-art-3vars}
  \begin{subfigure}[b]{.5\textwidth}
    \scriptsize
    \renewcommand{\arraystretch}{1.1}
    \begin{tabularx}{\textwidth}{lX}
      \hline
      \textbf{SIMULATION DETAILS} \\
      \hline
      \textbf{Agents}: Bidder 1, Bidder 2, Bidder 3, Auctioneer \\ 
      \parbox{.95\textwidth}{\textbf{Simulations Run: $7 \times 7 \times 7 = 343$}} \\
      \textbf{Speaking Order:} Auctioneer, Bidder 1, Auctioneer,\\\hspace{.75em} Bidder 2, Auctioneer, Bidder 3, ... repeat\\
      \hline
      \textbf{VARIABLE INFORMATION}  \\
      \hline
      \textbf{Final price} \\ \hspace{.75em}\parbox{.95\textwidth}{\textbf{Measurement Question:} Auctioneer: ``What was the final bid for the piece of art at the end of the auction?''} \\ \hspace{.75em}\parbox{.95\textwidth}{\textbf{Variable Type:} Continuous} \\
      \hline
      \textbf{Bidder 1's maximum budget} \\ \hspace{.75em}\parbox{.95\textwidth}{\textbf{Attribute Treatments:} [`\$50', `\$100', `\$150', `\$200', `\$250', `\$300', `\$350']} \\ 
      \hspace{.75em}\parbox{.95\textwidth}{\textbf{Proxy Attribute:} Your max budget for the art} \\ 
      \hspace{.75em}\parbox{.95\textwidth}{\textbf{Variable Type:} Continuous} \\
      \hline
      \textbf{Bidder 2's maximum budget} \\ \hspace{.75em}\parbox{.95\textwidth}{\textbf{Attribute Treatments:} [`\$50', `\$100', `\$150', `\$200', `\$250', `\$300', `\$350']} \\
      \hspace{.75em}\parbox{.95\textwidth}{\textbf{Proxy Attribute:} Your max budget for the art} \\
      \hspace{.75em}\parbox{.95\textwidth}{\textbf{Variable Type:} Continuous} \\
      \hline
      \textbf{Bidder 3's maximum budget} \\ \hspace{.75em}\parbox{.95\textwidth}{\textbf{Attribute Treatments:} [`\$50', `\$100', `\$150', `\$200', `\$250', `\$300', `\$350']} \\
      \hspace{.75em}\parbox{.95\textwidth}{\textbf{Proxy Attribute:} Your max budget for the art} \\
      \hspace{.75em}\parbox{.95\textwidth}{\textbf{Variable Type:} Continuous} \\
      \hline \\
    \end{tabularx}
    \caption{Information for experimental design}
    \label{tab:auction-art-3vars-table}
\end{subfigure}
\begin{subfigure}[b]{.45\textwidth}
  \centering
  \begin{tikzpicture}[->,>=stealth',shorten >=1pt,auto,node distance=3cm,
                      thick,
                      main node/.style={circle, draw, font=\scriptsize, text width=1.8cm, align=center, minimum size=2cm}]

    \node[main node] (A) at (0,0) {Final Price \\ $\mu =\MeanFinalPrice$ \\ $\sigma^2 = \VarFinalPrice$};
    \node[main node] (B) at (-3,3.5) {Bidder 1 Budget};
    \node[main node] (C) at (-4.5,0) {Bidder 2 Budget};
    \node[main node] (D) at (-3,-3.5) {Budder 3 Budget};

   \path[every node/.style={font=\scriptsize}]
      (B) edge[red] node[align=center, text=red] {\BetaBidOne\\ \phantom{-}(\SEBidTwo)} (A)
      (C) edge[red] node[align=center, text=red] {\BetaBidTwo\\(\SEBidTwo)}(A)
      (D) edge[red] node[align=center, text=red] {\BetaBidThree \\\phantom{-}(\SEBidThree)} (A);
  \end{tikzpicture}
  \caption{Fitted SCM}
  \label{fig:auction-art-3vars-scm}
\end{subfigure}
  \begin{minipage}{\textwidth}
    \begin{footnotesize}
    \emph{Notes:
    Figure~\ref{tab:auction-art-3vars-table} provides the information automatically generated by the system to execute the experiment for the proposed hypothesis.
    Figure~\ref{fig:auction-art-3vars-scm} shows the fitted SCM from the experiment.
    }
    \end{footnotesize}
    \end{minipage}
\end{figure}
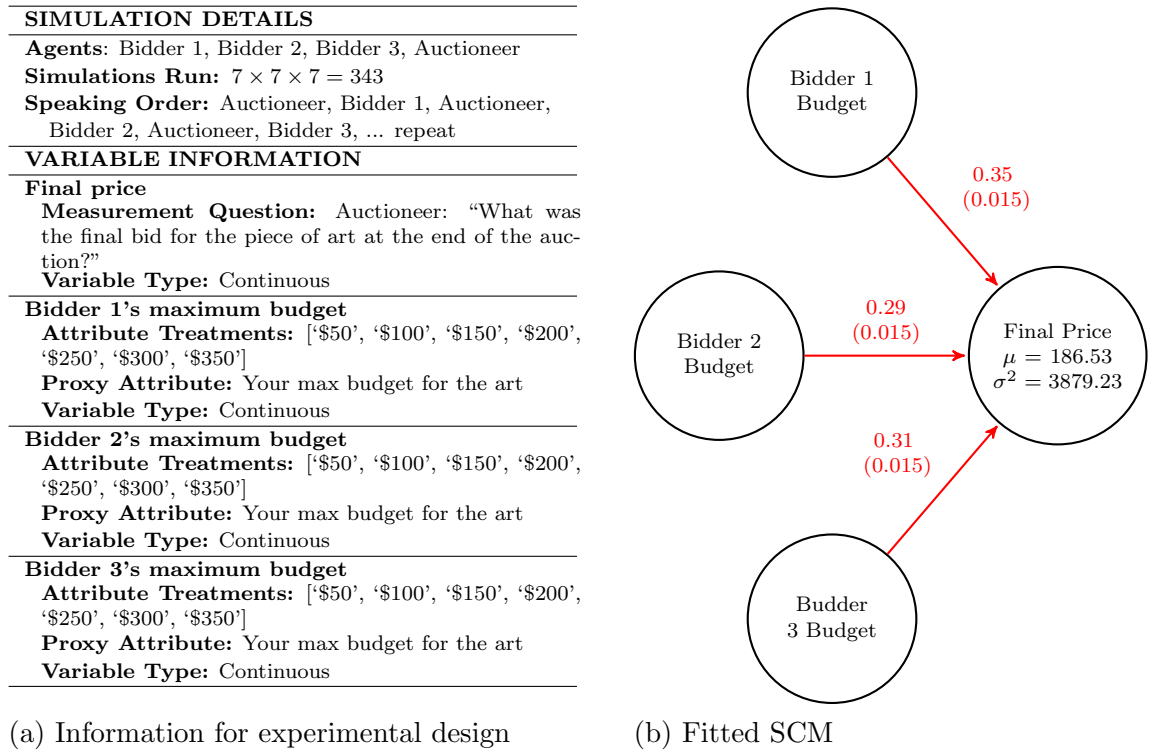

Finally, we explored the scenario of ``3 bidders participating in an auction for a piece of art starting at fifty dollars.''
Table \ref{tab:auction-art-3vars-table} shows that the causes are each bidder's maximum budget for the piece of art, and the outcome is the final price of the piece of art---all of which we selected.

All four variables are operationalized in dollars.
To maintain symmetry in the simulations, we also manually selected the same proxy attribute for the three bidders: ``your maximum budget for the piece of art.''
Each bidder had the same seven possible values for their attribute, leading to $7^3 = 343$ simulations of the auction.
It is important to note that these budgets are private values.
Unless a bidder publically reveals their budget, the other bidders do not know what it is.

Like the tax fraud scenario, the system chose the center-ordered interaction protocol for these simulations.
The auctioneer was selected as the central agent, and the other agents were bidder 1, bidder 2, and bidder 3, who alternated with the auctioneer in that order.

Figure \ref{fig:auction-art-3vars-scm} provides the results.
All three causal variables had a positive and statistically significant effect on the final price.
A one-dollar increase in any of the bidder's budgets caused a  \$\BetaWrittenBidOne, \$\BetaWrittenBidTwo, and \$\BetaBidThree \space increase in the final price for the piece of art for each respective bidder ($\hat{\beta}\text{*}  = \cdBidOne$, $\BidOnePvalString$; $\hat{\beta}\text{*} =\cdBidTwo$, $\BidTwoPvalString$; $\hat{\beta}\text{*} =\cdBidThree$ $\BidThreePvalString$).
These quantities make sense as each bidder has a $\frac{1}{3}$ chance of being marginal.

\section{LLM predictions for paths and points} \label{sec:predictions}

It is worth reiterating that the results in the previous section were not generated by directly prompting an LLM but rather through experimentation.
Although the experiments were fast and inexpensive, they were not free--in total, they took about 5 hours to run and cost over \$1,000.
This raises the question of whether the simulations were even necessary.
Could an LLM do a ``thought experiment'' (i.e., make a prediction based on a prompt) about a proposed \emph{in silico} experiment and achieve the same insight?
If so, we should just prompt the LLM to come up with an SCM and elicit its predictions about the relationships between the variables. 

To test this idea, we describe some of the simulations to the LLM and ask it to predict the results---path estimates and point predictions.\footnote{
  All predictions are made by the LLM once at temperature 0.
  When we elicit these predictions many times at higher temperatures, the results are similar.
}
Specifically, we modeled each scenario as $y = X \beta $, where $y$ is an $n \times 1$ vector and $X$ is a $n \times k$ matrix.
Here, $n$ is the number of simulations, and $k$ is the number of proposed causes.
The experiments from Section \ref{sec:results} provided us with estimates for $\hat{\beta}$ (a $k \times 1$ vector).
We describe the scenario and the experiment to the LLM and ask it to independently predict $y_i$ given each $X_i$ (a predict-$y_i$ task) as well as to predict $\hat{\beta}$ (a predict-$\hat{\beta}$ task).

The LLM's $y_i$ predictions are highly inaccurate compared to those from auction theory, which predicts that the clearing price will be the second highest valuation in an open-ascending price auction with private values \citep{PrivateAuctionMaskin1985}.
The LLM is also unable to accurately predict the path estimates ($\hat{\beta}$) of the fitted SCM.
Finally, we examine how the LLM does on the predict-$y_i$ task when provided with an SCM fit on all of the data except for the corresponding $X_i$ (the predict-$y_i|\hat{\beta}_{-i}$ task).
While the additional information dramatically improves the LLM's predictions, they are still less accurate than those made by auction theory.

\subsection{Predicting $y_i$}

For various bidder reservation price combinations in the auction experiment, we supply the LLM with a prompt detailing the simulation and experimental design.\footnote{
  In 80/343 simulations, the agents made the maximum number of statements (20) allowed by the system before the auction ended.
  We remove these observations because, without additional information, auction theory does not make predictions about partially completed auctions.
}
We then ask the LLM to predict the clearing price for the auction.
This gives us a point prediction for each simulated auction (i.e., each unique row $X_i$ in $X$) used to generate the fitted SCM in Figure~\ref{fig:auction-art-3vars-scm}.

\begin{figure}[htbp!]
  \caption{Comparison of the LLM's predictions to the theoretical predictions and a subset of experimental results for the auction scenario.}
  \label{fig:GPT-SCM}
  \centering
  \includegraphics[width=\linewidth]{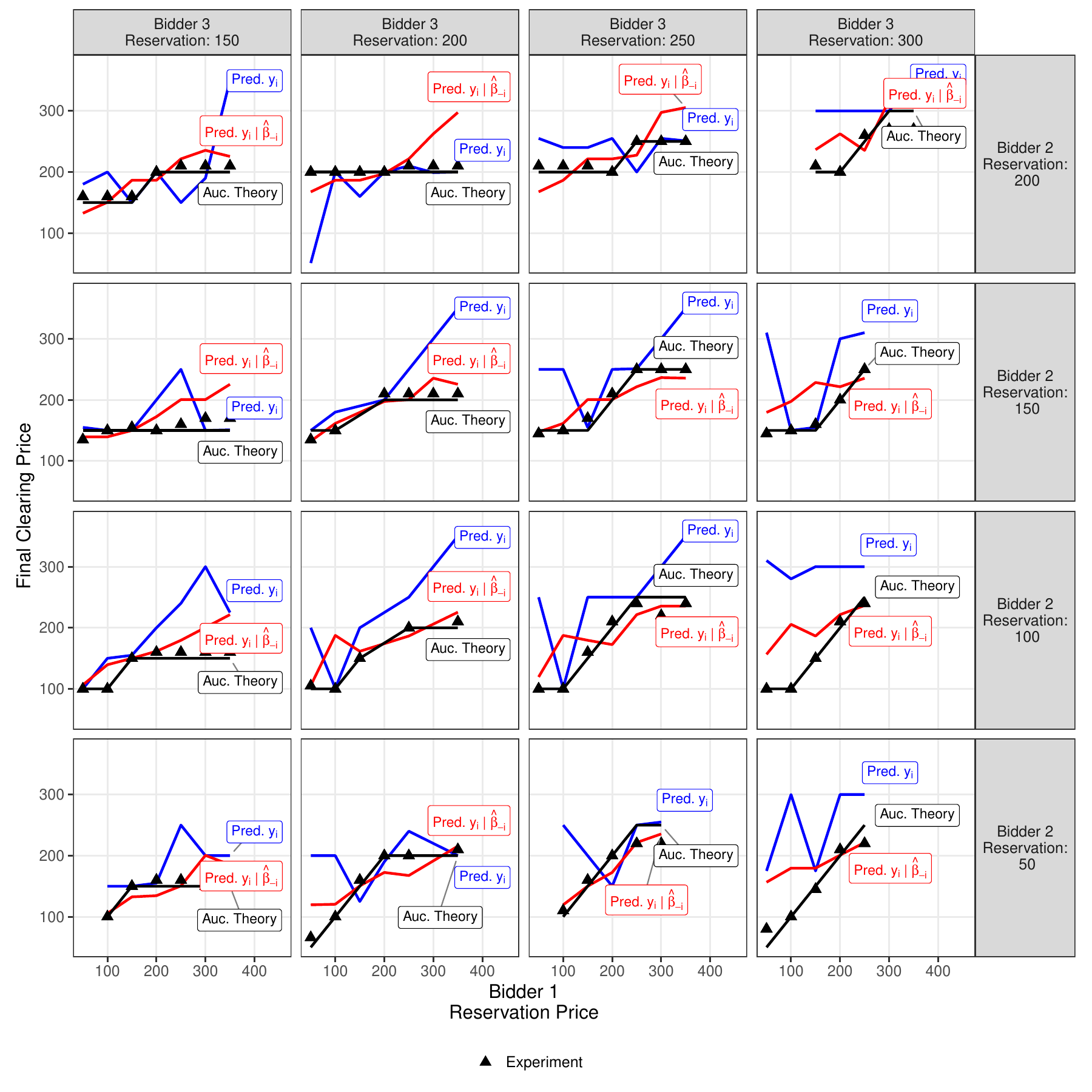}
  \begin{minipage}{\textwidth}
    \begin{footnotesize}
    \emph{Notes: The columns correspond to the different reservation values for bidder 3 in a given simulation, and the rows correspond to the different reservation values for bidder 2.
    The y-axis is the clearing price, and the x-axis lists bidder 1's reservation price.
    The black triangles track the observed clearing price in each simulated experiment, the black line shows the predictions made by auction theory ($MSE_{Theory} = \MSETheory$), the blue line indicates the LLM's predictions without the fitted SCM---the predict-$y_i$ task ($MSE_{y_i} = \MSEwoSCM$), and the red line is the LLM's predictions with the fitted SCM---the predict-$y_i|\hat{\beta}_{-i}$ task ($MSE_{y_i|\hat{\beta}_{-i}} = \MSEwSCM$).
}
    \end{footnotesize}
    \end{minipage}
\end{figure}

Figure \ref{fig:GPT-SCM} presents a comparison of the LLMs predictions, the simulated experiments, and the predictions made by auction theory.\footnote{
  We provide only a subset of the results in the main text as it is difficult to visualize all of them in a single figure.
  Figure~\ref{fig:GPT-SCM-full} shows the full set of predictions. The results are generally the same.
}
The columns correspond to the different reservation values for bidder 3 in a given simulation, and the rows correspond to the different reservation values for bidder 2.
The y-axis is the final bid price, and the x-axis lists bidder 1's reservation price.
The black triangles track the observed clearing price in each simulated experiment, the black line shows the predictions made by auction theory, and the blue line indicates the LLM's predictions without the fitted SCM---the predict-$y_i$ task.

The LLM performs poorly at the predict-$y_i$ task.
The blue line is often far from the black triangles and sometimes remains constant or even decreases as the second-highest reservation price across the agents increases.
In contrast, auction theory is highly accurate in its predictions of the final bid price in the experiment---the black line often perfectly tracks the black triangles.\footnote{
There are a few observations where the empirical clearing price is slightly above or below the theory prediction.
 In most cases where it was off, this was due to the auctioneer incrementing the bid price above the second-highest reservation price in the last round.
 }
The mean squared error (MSE) of the LLM's predictions in the predict-$y_i$ task ($MSE_{y_i} = \MSEwoSCM$) is an order of magnitude higher than that of the theoretical predictions ($MSE_{Theory} = \MSETheory$), and the predictions are even further from the theory than they are from the empirical results ($MSE_{y_i-Theory} = \MSEwoSCMTheory$).\footnote{
  MSE is reported for all predictions, not just the subset shown in Figure~\ref{fig:GPT-SCM}.
}

\subsection{Predicting $\hat{\beta}$}

We prompted the LLM to predict the path estimates and whether they would be statistically significant for the simulated experiments in Section \ref{sec:results}.
This is the predict-$\hat{\beta}$ task.
We then compare the LLM's predictions to the fitted SCMs.
With four experiments and three causes in each, we generate 12 predictions.

We provide the LLM with extensive information to make its predictions for each experiment.\footnote{See Figure~\ref{fig:beta-predict} in the appendix for the full prompt.}
This information includes the proposed SCM, the operationalizations of the variables, the number of simulations, and the possible treatment values.
Each prediction is elicited once at temperature 0.

The predictions are shown in Table \ref{tab:predictions}.
They were, on average, \MeanMagPred \space times larger than the actual estimates, and 10/12 of the predictions were overestimates.
Even when we remove the largest overestimate, the average magnitude of the ratio between the predicted and actual estimates is still \MeanMagNoOutlier.
The sign of the estimate was correct in 10/12 predictions, and 10/12 correctly guessed whether or not the estimate would be statistically significant.
When we repeat the predictions at a higher temperature and take their average, the results are similar (see Table \ref{tab:predictions_1}).

\subsection{Predicting $y_i|\hat{\beta}_{-i}$} 

The LLM was, on average, off by an order of magnitude for both the predict-$y_i$ task and the predict-$\hat{\beta}$ task, but maybe it can do better with more information.
For each $X_i$ in the auction simulations, we use the data from the experiment to estimate $\hat{\beta}_{-i}$, the path estimates from the SCM excluding the $i$th observation.
We then prompt the LLM to predict the outcome for each $X_i$ given $\hat{\beta}_{-i}$.

The red line in Figure \ref{fig:GPT-SCM} provides these new predictions.
The LLM's predictions are much closer to the actual outcomes when it has access to a fitted SCM ($MSE_{y_i|\hat{\beta}_{-i}}= \MSEwSCM$) as opposed to when it does not ($MSE_{y_i} = \MSEwoSCM$), even though all the predictions are out of sample and every $X_i$ is unique.

However, the LLM's predictions on the predict-$y_i|\hat{\beta}_{-i}$ task are still not as accurate as the predictions made by auction theory
($MSE_{Theory} = \MSETheory$).\footnote{
  It is also less accurate than the mechanical predictions made by the fitted SCM using the same procedure $MSE_{Mechanistic: y_i|\hat{\beta}_{-i}} = \MSEMechanistic$.
  Maybe the LLM cannot do the math, is still conditioning on other information beyond the path estimates when making its predictions, or, like humans, is ignoring relevant information when making choices \citep{NoUseInfo2018Handel}.
}
They are also still further from the theory than they are from the empirical results ($MSE_{y_i|\hat{\beta}_{-i}-Theory} = \MSEwSCMTheory$).
There is clearly room for improvement.
That improvement is feasible with the system: there exists an SCM perfectly consistent with auction theory.
Only one exogenous variable was missing: the second-highest reservation price of the bidders.
If allowed to generate and test enough potential causes, our system could have selected this variable as a possible cause by itself.
In this case, the fitted SCM would have matched the theoretical predictions.\footnote{
  When we do fit this SCM (see Figure~\ref{fig:auction-art-theory}), the coefficient is close to one ($\beta = \SCMTheorySimpleEst$), and almost all the variance in the outcome is explained ($R^2 = \SCMTheorySimpleRsq$).
}
\section{Identifying causal structure ex-ante}  \label{sec:advantages}

The SCM-based approach offers a promising new method for studying simulated behavior at scale.
However, it is not the only option for such rapid exploration.
Others have designed large, quasi-unstructured simulations demonstrating exciting results.
For example, \cite{park2023generative} endows a group of LLM agents with personas and memory systems and then allows them to freely interact in a simulated community for an extended period.
Despite no explicit instructions to do so, the agents in the simulation produce many human-like behaviors, such as throwing parties, going on dates, and making friends.

While impressive and informative, a problem with such open-ended social simulations is that selecting and analyzing outcomes can be difficult.
To unveil insights, researchers may need to comb through thousands of lines of unstructured text.
If they are interested in casual relationships, they may need to infer the causal structure ex-post, which can be problematic.
In contrast, the SCM framework describes exactly what needs to be measured as a downstream outcome subject to the exogenous manipulations of the cause.
Identification is guaranteed.
In this section, we discuss how assuming or searching for causal structure in observational data, the type generated from massive open-ended simulations can lead to misidentification and how using SCMs avoids this problem.

\subsection{Assuming causal structure from data}

All estimates in the fitted SCMs in Section \ref{sec:results} are unbiased.
We know this because the data comes from an experiment, and we randomized on the causal variables.
A nice feature of a perfectly randomized experiment is that we can get unbiased measurements of any downstream endogenous outcome relative to the exogenous manipulations.\footnote{
	When we say ``downstream,'' we mean any variable whose value is realized after the agents begin interacting in the simulated conversations.
}
I.e., the coefficients on the fitted SCM are identified.
For example, in the bargaining experiment, perhaps we are interested in the length of the conversation as an outcome, even though it was not a part of the original SCM.
The conversation length can be operationalized as the sum of the number of statements made by all agents, and we can use the transcript from the finished experiment to measure it.
We can then fit an SCM with the data and get unbiased estimates of the effect of the exogenous variables on the conversation's length.

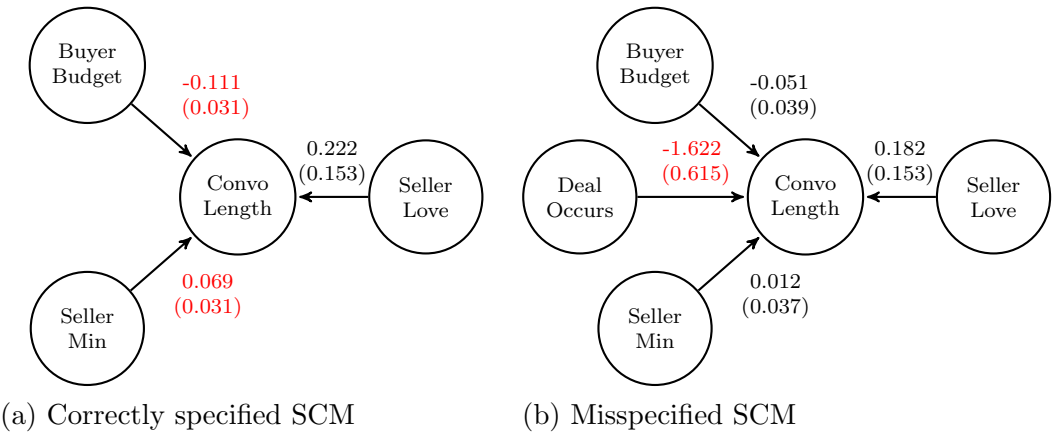
\begin{figure}[htbp]
  \caption{Comparison of the true and misspecified SCMs.}
  \label{fig:compare-specification}
  \centering
  \begin{subfigure}{.45\textwidth}
    \centering
    \begin{tikzpicture}[->,>=stealth',shorten >=1pt,auto,node distance=3cm,
                        thick,
                        main node/.style={circle, draw, font=\scriptsize, text width=1cm, align=center, minimum size=1cm}]

      \node[main node] (A) at (2,-1.5) {Convo Length};
      \node[main node] (B) at (0,.25) {Buyer Budget};
      \node[main node] (C) at (0,-3.25) {Seller Min};
      \node[main node] (D) at (4.5,-1.5) {Seller Love};

     \path[every node/.style={font=\scriptsize}]
        (B) edge node[align=center, color = red] {\TrueBudgetBeta\\ \phantom{-}(\TrueBudgetSe)} (A)
        (C) edge node[align=center, swap, color = red] {\TrueSellerMinBeta\\ (\TrueSellerMinSe)} (A)
        (D) edge node[align=center, swap] {\TrueLoveBeta\\ (\TrueLoveSe)} (A);
    \end{tikzpicture}
    \caption{Correctly specified SCM}
    \label{fig:scm-specified}
\end{subfigure}
\begin{subfigure}{.45\textwidth}
  \centering
  \begin{tikzpicture}[->,>=stealth',shorten >=1pt,auto,node distance=3cm,
                      thick,
                      main node/.style={circle, draw, font=\scriptsize, text width=1cm, align=center, minimum size=1cm}]
    % nodes
    \node[main node] (A) at (2,-1.5) {Convo Length};
    \node[main node] (D) at (-1,-1.5) {Deal Occurs};
    \node[main node] (B) at (0,.25) {Buyer Budget};
    \node[main node] (C) at (0,-3.25) {Seller Min};
    \node[main node] (E) at (4.5,-1.5) {Seller Love};
   % edges
   \path[every node/.style={font=\scriptsize}]
      (B) edge node[align=center] {\MissBudgetBeta\\\phantom{-}(\MissBudgetSe)} (A)
      (C) edge node[align=center, swap] {\MissSellerMinBeta\\(\MissSellerMinSe)}(A)
      (D) edge node[align=center, color = red] {\MissDealBeta\\\phantom{-}(\MissDealSe)} (A)
      (E) edge node[align=center, swap] {\MissLoveBeta\\(\MissLoveSe)} (A);
  \end{tikzpicture}
  \caption{Misspecified SCM}
  \label{fig:scm-misspecified}
\end{subfigure}
  \begin{minipage}{\textwidth}
    \begin{footnotesize}
    \emph{Notes: 
    Statistically significant paths are marked in red ($\alpha = 0.05$).
    Each path is given with its estimated coefficient and standard error in parentheses.
    Both SCMs are estimated using the data from the bargaining scenario in Section \ref{sec:results}.
    Subfigure (a) provides a correctly specified SCM from a randomized experiment.
    Subfigure (b) shows a misspecified SCM based on an assumed structure.
    The path estimates of the buyer's budget and the seller's minimum price go from significant in the correctly specified SCM to insignificant and far closer to zero in the misspecified SCM.
    }
    \end{footnotesize}
    \end{minipage}
\end{figure}

Figure \ref{fig:scm-specified} shows this fitted SCM using the data from the experiment in Section~\ref{sec:results}.
Both the buyer's budget and the seller's minimum price have a significant effect on the length of the conversation ($\TrueBudgetPvalString$; $\TrueSellerMinPvalString$), but the seller's emotional attachment does not ($\TrueLovePvalString$).

Suppose we did not know the actual causal structure of these scenarios or that the data came from an experiment.
All we have are the data for the original three causes, the conversation length, and whether a deal was made (the original outcome).
If we want to estimate the causal relationships between these variables, we would have to make untestable assumptions.
For example, one could reasonably presume that the buyer's budget, the seller's minimum price, the seller's emotional attachment, and whether a deal was made all causally affect the length of the conversation.

Figure \ref{fig:scm-misspecified} provides the fitted SCM for this proposed causal structure.
Only whether a deal was made was estimated to have a significant effect on the length of the conversation ($\MissDealPvalString$).
But we know this is wrong. 
We have the true causal structure in Figure \ref{fig:scm-specified} from a perfectly randomized experiment, and both the buyer's and the seller's reservation prices had a significant effect on the length of the conversation.
Here, they are insignificant and far closer to zero ($\MissBudgetPvalString$; $\MissSellerMinPvalString$).
Whether or not the deal occurred is a bad control that biases the estimates---it is probably codetermined with the length of the conversation.\footnote{
  We cannot be sure about the causal relationship between the length of the conversation and whether a deal was made because neither is exogenously varied in the experiment.
  All we know is that controlling for whether or not a deal occurs induces bias, as we have the experiment as a reference.
}

The informed econometrician may presume that she would never make such a mistake, but many researchers are not so savvy.\footnote{
LLMs are definitely not yet savvy enough to avoid this mistake.
}
We were unsure of it until we had unbiased estimates from the correctly specified SCM as a reference.
There are also many kinds of bad controls, and many of them are less obvious than those in this example \citep{cinelli2022crash}.
It is easy to misspecify a model when the data is observational and has many variables, even when their relationships may seem obvious.

The SCM-based approach avoids the bad controls.
The generation of the data is based on the causal structure.
There is no need to instrument endogenous variables and presume their causal relationships.
Exogenous variation is explicitly induced in the SCM to identify the causal relationships ex-ante.
Even if we do not know how a new outcome is incorporated into the causal structure, we can always reference how it is affected by the exogenous variables by fitting a simple linear SCM.
 
\subsection{Searching for causal structure in data}

Another strategy for identifying causal relationships when the underlying structure is unknown is to let the data speak for itself. 
For example, we could use an algorithm to find the model that makes the data most likely.
There are many ways to do this, none of which can always, or even consistently, identify the correct causal relationships from observational data \citep{pearl2009causal}.
These algorithms take as input potential variables of interest (a graph with no edges, only nodes) and data for these variables.
They output a proposed DAG that best fits the data.\footnote{
  These algorithms often do not presume a functional form, so we refer refer to hypotheses as DAGs, not SCMs, in this section.
}

The simplest algorithm is to generate all possible DAGs for existing variables and then evaluate each model based on some criteria (e.g., maximum likelihood, Bayesian information criterion, etc.).\footnote{
  The number of possible DAGs grows exponentially with the number of nodes. For example, for $n=1,2,3,$ and 4 nodes, there are 1, 3, 25, and 543 possible DAGs.
  This is a combinatorial explosion, and it is not feasible to evaluate all potential models for a large number of nodes, which presents further problems for this approach.
  }
Another method is to add edges that maximize the criteria greedily.
This approach can be further improved by penalizing the model for complexity (based on additional criteria) and removing edges until the model is greedily optimized.
The second approach is the Greedy Equivalence Search (GES) algorithm \citep{chickering2002optimal}, which we used on the data and from all the experiments in Section \ref{sec:results}.\footnote{
  The GES algorithm is not perfectly stable; different runs on the same data can produce different results, which is its own problem.
}

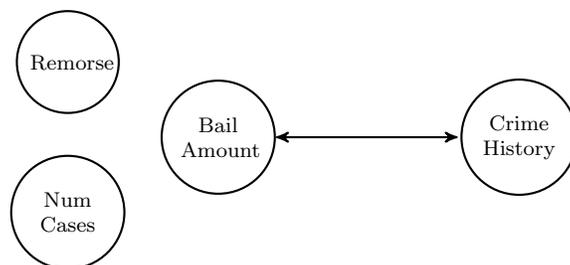
\begin{figure}[htbp]
  \caption{Incorrect causal structure identified by the GES algorithm for the tax fraud experiment.}
  \label{fig:scm-search}
  \centering
    \centering
    \begin{tikzpicture}[->,>=stealth',shorten >=1pt,auto,node distance=3cm,
                        thick,
                        main node/.style={circle, draw, font=\scriptsize, text width=1cm, align=center, minimum size=1cm}]
      % nodes
      \node[main node] (A) at (-2, 0) {Bail Amount};
      \node[main node] (D) at (2, 0) {Crime History};
      \node[main node] at (-4, 1) (B) {Remorse};
      \node[main node] at (-4, -1) (C) {Num Cases}; 
     % edges
     \path[every node/.style={font=\scriptsize}]
    (A) edge[<->] node[align=center] {} (D);
    \end{tikzpicture}
  \begin{minipage}{\textwidth}
    \begin{footnotesize}
    \emph{Notes: The Greedy Equivalence Search (GES) algorithm can incorrectly identify the causal structure of observational data.
    In the tax fraud scenario, we know from Figure \ref{fig:tax-fraud-scm} and the accompanying experiment that an increase in the defendant's previous convictions caused an increase in the average bail amount.
    However, the algorithm identified the causal relationship as equally likely in either direction.
    Without the correctly specified DAG, a researcher would have to assume the causal structure of the data, which can be problematic.
    }
    \end{footnotesize}
    \end{minipage}
\end{figure}

In some experiments, the algorithm incorrectly identified the causal structure.
Figure~\ref{fig:scm-search} provides the DAG identified by the GES algorithm for the tax fraud scenario.
As a reminder, the original causal variables are the defendant's previous convictions, the judge's number of cases heard that day, and the defendant's level of remorse, and the outcome is the bail amount. 
The algorithm has no information about which variables are exogenously varied, just the raw data.

The GES algorithm identified the defendant's criminal history and the bail amount as the only variables in the scenario with any causal relationship.
This is partially correct---we know from the experiment that an increase in the defendant's previous convictions caused an increase in the average bail amount.
However, the algorithm identified the causal relationship as equally likely in either direction.
There was no more evidence in the data that the defendant's criminal history caused the bail amount than the bail amount caused the defendant's criminal history.
And while we know that the former is correct from our experiment, a researcher using the algorithm without the correctly specified DAG would not.
They would have to make an assumption, which, as we have shown, can be problematic.

The SCM-based approach avoids search problems, as we never need to search for the causal structure given the data.
Instead, we generate the data based on a proposed causal structure.
Even if we want to measure a new outcome on the existing experimental data, we have already identified the sources of exogenous variation.

We should note that problems with searching for or assuming causal structures from data are not new.
\cite{pearl2009causal} makes a similar point many times.
However, social scientists have never had the tools to induce exogenous variation and explore causal relationships at scale in many different scenarios.

\section{Conclusion} \label{sec:conclusion}

This paper demonstrates an approach to automated \emph{in silico} hypothesis generation and testing made possible through the use of SCMs.
We implemented the approach by building a computational system with LLMs and provided evidence that simulations can elicit information from an LLM that was not ex-ante available to the model.
We also showed that such simulations produce results that are highly consistent with theoretical predictions made by the relevant economic theory.
In this final section, we will discuss why such systems could be useful and identify areas for future research.

\subsection{Controlled experimentation at scale}

How might systems like the one presented in this paper be useful for social science research?
One view is that these simulations are simple dress rehearsals for ``real'' social science.
A more expansive and exciting view is that these simulations would yield insights that sometimes generalize to the real world.

This is a view that sees these agents as a step forward in representing humans far beyond classical methods in agent-based modeling, such as those used to explore how individual preferences can lead to surprising social patterns \citep{schelling1971dynamic, schelling1969models}.\footnote{
	See \cite{horton2023large} for a full discussion on the differences between traditional agent-based modeling and the use of LLM-powered agents.
  This position reflects our views as it was written recently by some of the authors of this paper.
}
This view would mirror recent advances in the use of machine learning for protein folding \citep{jumper2021highly} and material discovery \citep{merchant2023scaling}.

The system presented in this paper can generate these controlled experimental simulations en masse with prespecified plans for data collection and analysis.
That contrasts most academic social science research as currently practiced \citep{20questions2022}.\footnote{
  When a group of social scientists has the same data set on some human behavior or outcome, they can reach very different conclusions when analyzing it independently \citep{fragilefamily2020, uncertainuniverse2023}.
}
This contrast is important.
In the social sciences, context can heavily influence results.
Outcomes that hold true for one population may not for another.
Even within the same population, a change in environment can nullify or flip results \citep{Endow2004Lerner}.
Studying humans is also expensive and time-consuming, which makes rapid, inexpensive, and replicable exploration valuable.
There is still, of course, the fundamental jump from simulations to human subjects.

\subsection{Interactivity}

The system allows a scientist to monitor its entire process.
Should a researcher disagree with or be uncertain about a decision made by the system, they can probe the system regarding its choice.
This allows the researcher to either (1) understand why the decision was made, (2) ask the system to come up with a different option for that decision, or (3) input their own custom choice for that decision. 

A researcher can even ignore much of the automation process and fill in the details themselves.
They can choose the variables of interest, their operationalizations, the attributes of the agents, how the agents interact, or customize the statistical analysis, among other decision points.
Different parts of the system can also accommodate different types of LLMs simultaneously.
For example, a researcher could use GPT-4 to generate hypotheses and Llama-2-70B to power the agents' simulated social interactions.

\subsection{Replicability}

Replicating social science experiments with human subjects can be difficult \citep{camerer2018replicate}.
Despite the use of preregistrations, the exact procedures used in experiments are often unclear \citep{uncertainuniverse2023}.
In contrast, the system allows for nearly frictionless communication and replication of the experimental design.

The system's entire procedure is exportable as a JSON file with the fitted SCM.\footnote{
  A JSON (JavaScript Object Notation) is a data format that is easy for humans to read and write and easy for machines to parse and generate. It is commonly used for transmitting data in web applications, as a configuration and data storage format, and for serializing and transmitting structured data over a network.
}
This JSON includes every decision the system makes, including natural language explanations for the choices and the transcripts from each simulation.
These JSONs can be saved or uploaded at any point in the system's process.
A researcher could run experiments and post the JSON and results online.
Other scientists could inspect, replicate the experiment, or extend the work.

\subsection{Future research}

While designing our system, we encountered several areas for new research.
First is the problem of ``which attributes'' to endow an LLM-powered agent beyond those immediately relevant to the proposed exogenous variables. 
For example, demographic information, personalities, and other traits are not included in the agent's attributes unless they are a part of the SCM. 
To improve the fidelity of the simulations, it might make sense to add some or all of these attributes to the agents.
However, it is unclear how to optimize this process.

Second, we encountered the problem of engineering social interactions between LLM agents. 
LLMs are designed to exchange text in sequence, necessitating a protocol for turn-taking that reflects the natural ebb and flow of human conversation. 
In an initial attempt to address this problem, we created a menu of flexible agent-ordering mechanisms. 
We also introduced an additional LLM-powered agent into our version of the system whom we dub the `coordinator.''
The coordinator functions as a quasi-omniscient assistant who can read through transcripts and make choices about the speaking order of other agents in the simulations.
There are probably better ways to determine the speaking order of agents.

A related problem is the question of when to stop the simulations.
Like Turing's halting problem, there is likely no universal rule for when conversations should end, but there are probably better rules than those we have implemented.
A Markov model approximating the distribution of agents speaking, estimated from real conversation data, might provide more naturalistic results for simulating and ending interactions, but that is an idea for future work.

Lastly, if we can build a system that can automate one iteration of the scientific process and determine a follow-on experiment, a clear next step is to set up an intelligently automated research program.
This would involve using outcomes from the simulations to inform continuous cycles of experimentation.
Then, a researcher could intelligently explore a given scenario's parameter space.
How to optimize this exploration amongst so many possible variables will be an important problem to solve.

As presented in this paper, the system provides only one possible implementation of the SCM-based approach.
We made many subjective decisions.
Other researchers might implement the approach with different design choices.
There is room for improvement and exploration.

\newpage \clearpage
\bibliographystyle{aer}
\bibliography{rs}

\newpage \clearpage
\appendix

\setcounter{figure}{0}
\renewcommand{\thefigure}{A.\arabic{figure}}
\setcounter{table}{0}
\renewcommand{\thetable}{A.\arabic{table}}

\section{Implementation details} \label{sec:implementation}

The first step in the system's process is to query an LLM for the roles of the relevant agents in the scenario.
When we say ``query an LLM,'' we mean this quite literally.
We have written a scenario-neutral prompt that the system provides to an LLM with the scenario added to the prompt.
The prompt is scenario-neutral because we can reuse it for any scenario.
The prompt takes the following format:
\begin{quote}
In the following scenario: ``\{\texttt{scenario\_description}\}'',
Who are the individual human agents in a simple simulation of this scenario?
\end{quote}
where \{\texttt{scenario\_description}\} is replaced with the scenario of interest.
The LLM then returns a list of agents relevant to the scenario, and we have various checking mechanisms to ensure the LLM's response is valid.

The system contains over 50 pre-written scenario-neutral prompts to gather all the information needed to generate the SCM, run the experiment, and analyze the results.
These prompts have placeholders for the necessary information aggregated in the system's memory as it progresses through the different parts of the process.

\subsection{Constructing variables and drawing causal paths}

The system builds SCMs variable-by-variable.
It queries an LLM for an outcome involving the agents in the social scenario of interest.
We refer to outcomes as endogenous variables because their values are realized during the experiment. 
This contrasts exogenous variables, the causes, whose values are determined before the experiment.

The system queries the LLM for a list of possible exogenous causes of the endogenous variable, generating a hypothesis as an SCM.\footnote{
  There is growing evidence that LLMs can be quite good at coming up with ideas and generating hypotheses \citep{rosenbuschHyp2023,girotra2023ideas}.
}
Exogenous variables serve as inputs to the experiment, whose values can be deterministically manipulated to identify causal effects.
The system assumes that when an exogenous variable causes an endogenous variable, a single causal path is proposed from the exogenous variable to the endogenous variable.
More formally, the system always generates SCMs as a simple linear model.
The system currently generates all SCMs with one endogenous variable and as many exogenous causes as a researcher desires.
We do little optimization here, although the system can test for interaction terms.
In future iterations of the system, a researcher could choose outcomes and causes they are interested in, score hypotheses by interestingness, and generate more complex hypotheses with mediating endogenous variables.\footnote{
  Parallel and crossover experimental designs can be used to identify mediating causal relationships \citep{CauselMechanisms2012Imai}.
  These experiments require few assumptions, which are often more plausible when researchers have more control over the experiment, as they usually do with LLMs.
}

\subsubsection{Endogenous outcomes}

For each endogenous variable, the system generates an operationalization, a type, the units, the possible levels, the explicit questions that need to be asked to measure the variable's realized value, and how the answers to those questions will be aggregated to get the final data for analysis. 
Examples of all information collected about the variables in an SCM are provided in Table~\ref{tab:variable_information}.
Each piece of information about a variable is stored by the system and is then used to determine subsequent information in consecutive scenario-neutral prompts.
This is a kind of ``chain-of-thoughts prompting'', or the process of breaking down a complex prompt into a series of simpler prompts.
This method can dramatically improve the quality and robustness of an LLM's performance \citep{chain_thoughts:2023}.

The first piece of information determined for each endogenous variable is the operationalization.
That is, how the possible realizations of said variable can be directly mapped to measurable outcomes that can be observed and quantified.
Suppose the outcome variable is \texttt{whether or not a deal occurred} from the SCM in Figure~\ref{fig:mug-love-scm}.\footnote{
  We continue the practice from Section~\ref{sec:overview} of using \texttt{typewriter text} to denote example information from the system.
}
The system could operationalize this as a binary variable, where \texttt{``1'' means a deal occurred and ``0'' does not}.
It then stores this information and uses it in a scenario-neutral prompt to choose the variable type.

All variables are determined to be one of five mutually exclusive ``types.'' 
These are continuous, ordinal, nominal, binary, or count.
By selecting a unique type for each variable, the system can accommodate different distributions when estimating the fitted SCM after the experiment.

Each variable also has units.
The units are the specific measure or standard used to represent the variable's quantified value.
This information is used to improve the robustness and consistency of the system's output when querying the LLM for other information about a variable.

The levels of the variable represent all of the values the variable can realize in a short list.
They can take on different forms depending on the variable type, but they all follow a general pattern where they are defined by the range and nature of a variable's possible values.\footnote{
  For binary variables, the levels are the two possible outcomes.
  For ordinal variables, the levels include all possible values that the ordinal variable could take on as determined by its operationalization.
  The levels are selected for count and continuous variables by segmenting the range of possible values into discrete intervals. 
  In cases where the variable does not have a defined maximum or minimum, categories such as ``above X'' or ``below Y'' are included to ensure all possible values are covered.
}

To measure the endogenous outcome, the system generates survey questions for one of the agents.
For example, to measure \texttt{whether or not a deal occurred}, the system could ask the buyer or the seller, ``\texttt{Did you agree to buy the mug?}''
Or, if the endogenous variable was \texttt{the final price of the mug}, the system could ask one of the agents, ``\texttt{How much did you sell the mug for?}''
Even though the simulations have yet to be conducted, the system generates survey questions.
As with pre-registration, this reduces unneeded degrees of freedom in the data collection process after the experiment.

Most endogenous variables are measured with only one question.
In this case, the answer to this question is the only information needed to quantify the variable.
Sometimes, it takes more than one survey question to measure a variable.
Maybe the variable is the \texttt{average satisfaction of the buyer and the seller}; a variable that requires two separate measurements to quantify.
In this case, the system generates separate measurement questions to elicit the buyer's and the seller's satisfaction.
Then, the system averages the answers to the questions to measure the variable.

We pre-programmed a menu of 6 mechanical aggregation methods: finding the minimum, maximum, average, mode, median, or sum of a list of values.
If the system needs to combine the answers to multiple questions to measure a variable, it queries an LLM to select the appropriate aggregation method.
Then, the system uses a pre-written Python function to perform said aggregation.
We refrain from asking the LLM to perform mathematical functions whenever possible, as they often make mistakes.

\subsubsection{Exogenous causes}

Besides the explicit measurement questions and data aggregation method, the system collects the same information for the exogenous variables as it does for the endogenous variables.
For exogenous variables, these two pieces of information are unnecessary for measurement.
In each simulation of the social scenario, a different combination of the values of the exogenous variables is initialized.
This is how the system induces variation in an experiment, so the treatments are always known to the system ex-ante.

Causal variables can have one of two possible ``scopes.''
The scope can be specific to an individual agent or the scenario as a whole.
This scope determines how the system induces variation in the exogenous variables---at the agent or scenario level.
Individual-level variables are further designated as either public or private.
If private, the variable's values are only provided to one agent; if public, they are treated as common knowledge to all agents in the scenario.

The system induces variation in the exogenous variables by transforming them into manageable proxy attributes for the agents.
The system queries an LLM to create a second-person phrasing of the operationalized variable provided to the agent (or agents, depending on the scope).
For instance, with the \texttt{buyer's budget} variable, the attribute could be ``\texttt{your budget}'' for the buyer.
These attributes will be assigned to the agents, which we discuss in Section \ref{sec:agents}.

With the proxy attribute for the variable, the system queries an LLM for possible values the attribute can take on.
These are the induced variations---the treatment conditions for the simulated experiments.
By default, the system uses the levels, or a value within each level, of the variable for the possible variation values.
For example, these could be $\{\text{\$5, \$10, \$20, \$40}\}$ for the \texttt{buyer's budget}.

\subsection{Building hypothesis-driven agents}   \label{sec:agents}

In conventional social science research, human subjects are catch as catch can.
Here, we have to construct them from scratch.
By ``construct'' we mean that we prompt an LLM to be a person with a set of attributes.
This is quite literal; for example, we could construct an agent in a negotiating scenario with the following prompt:
\begin{quote}
  ``You are a buyer in a negotiation scenario with a seller.
  You are negotiating over a mug.
  You have a budget of \$20.''
\end{quote}
We can construct an agent with any set of attributes we want, which raises the question of what attributes we should use.

We already have the attributes that will be varied to test the SCM, but there are many others we could include.
Some work has explored the endowing of agents with many different attributes, but it is unclear what is optimal, sufficient, or even necessary.\footnote{
  The methods have varied, ranging from endowing agents with interesting attributes \citep{horton2023large, argyle2023out} to using American National Election Study data to create ``real'' people \citep{tornberg2023simulating} to demonstrating that endowing demographic information does not necessarily represent a population of interest \citep{Which_humans:2023, santurkar2023opinions}. 
  There is a balance to be struck. 
  While attributes can provide a rich and nuanced simulation, they can also lead to redundancy, inefficiency, and unexpected interactions.
  In contrast, too few attributes might result in an oversimplified and unrealistic portrayal of social interactions. 
}
We take a minimalist approach, endowing our agents with goals, constraints, roles, names, and any relevant proxy attributes for the exogenous variables.
In the future, we could integrate large numbers of diverse agents, perhaps constructed to be representative of some specific population.

\subsubsection{Assigning agents attributes}

\begin{figure}[ht]
\caption{Example agents generated by the system for ``two people bargaining over a mug''}      
\label{fig:example_agents}
    \centering
   \includegraphics[width=\linewidth]{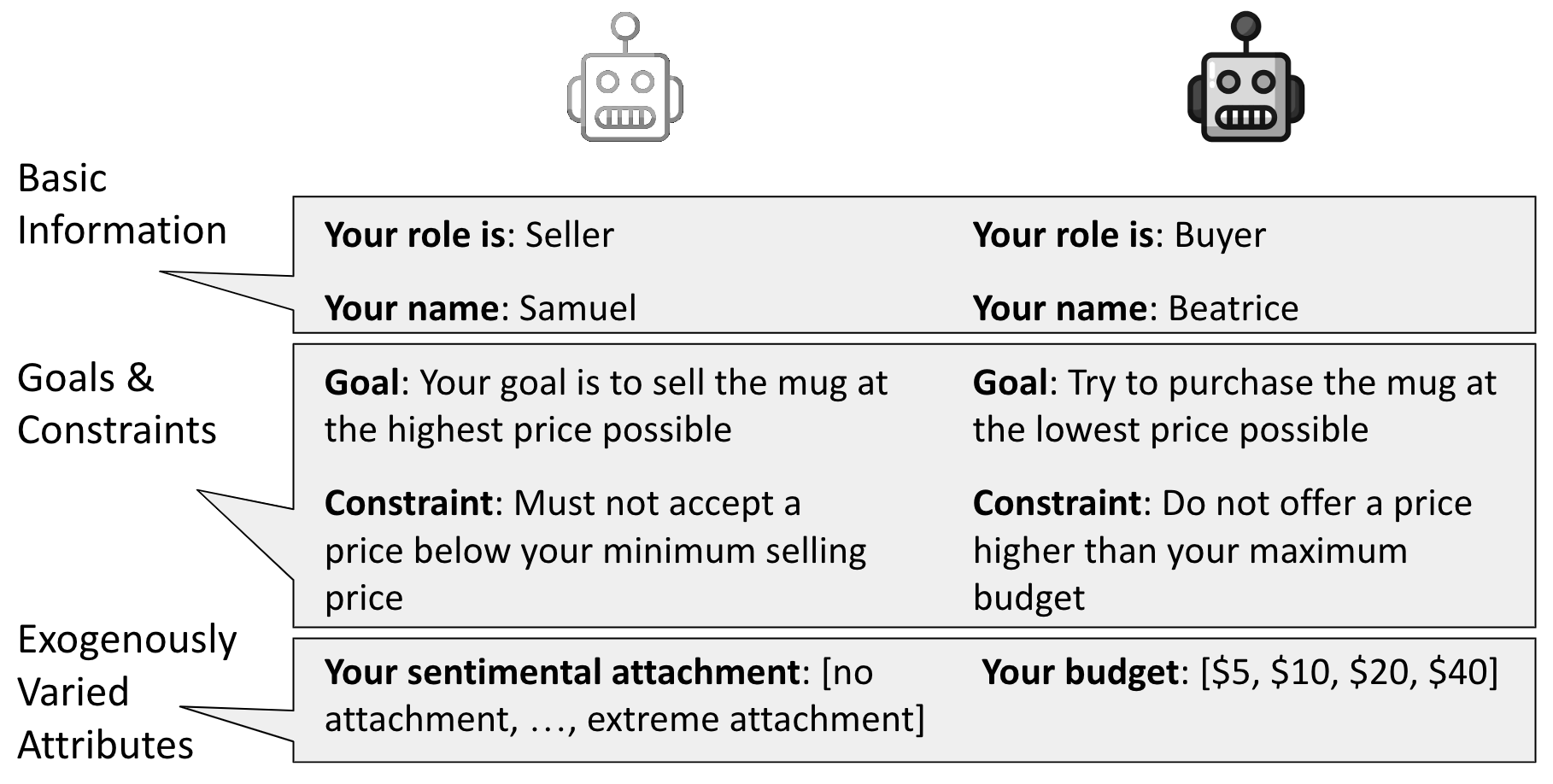}
   \begin{minipage}{\textwidth}
    \begin{footnotesize}
      \emph{Notes: 
      In all simulations, agents are endowed with a randomly generated name, role, goal, constraint, and proxy attributes for the exogenous variables.
      To simulate the experiment for the agents in this figure, the system will generate four versions of the seller and four versions of the buyer, each with one of the values for the exogenously varied attributes (assuming there are four possible values for ``Your sentimental attachment'').
      That is $4 \times 4 = 16$ treatments.}
    \end{footnotesize}
    \end{minipage}
\end{figure}

The system collects information for agents independently, similar to its one-at-a-time approach with the variables in the SCM.
The system randomly selects an agent, determines its attributes, and then moves on to the next agent.\footnote{
  The system already has the agent's roles from the construction of the SCM.
}
Examples of buyer and seller agents with their attributes are provided in Figure \ref{fig:example_agents}.

For each agent, the system queries the LLM for a random name.
Agents perform better in simulations with identifiers to address one another, although this feature can be disabled.
An agent's name can also be varied as a proposed exogenous cause.
The system then queries an LLM again, this time for a goal and then a constraint, which we discuss in the following subsection.

Finally, the system cross-checks the values of the proxy attributes between the agents to ensure they overlap appropriately.
For example, if the two exogenous variables in the SCM were the \texttt{buyer's budget} and the \texttt{seller's minimum acceptable price}, the system would check to make sure that \texttt{the seller's minimum acceptable price} is not invariably higher than \texttt{the buyer's budget}.
We let the LLM determine if these attribute values overlap appropriately. 
If any discrepancies are found, the system queries the LLM again to resolve them with new values for the proxy attributes.
Otherwise, the simulated experiment would waste time and resources because the induced variations were not supported across reasonable values.
For example, if the \texttt{buyer's budget} was always below the \texttt{seller's minimum acceptable price}, then they might never make a deal.

\subsubsection{The importance of agent goals}

Unlike, say, economic agents, whose goals are expressed via explicit utility functions, the LLM agent's goals are expressed in natural language. 
In the context of our bargaining scenario, an example goal generated by our system for the seller is to \texttt{sell the mug at the highest price possible}.
An example constraint is to \texttt{not accept a price below your minimum selling price.}
These goals and constraints are oriented towards value, but they do not have to be; these are merely the ones generated by the system.
A constraint could just have easily been \texttt{do not ruin your reputation with your negotiating partner}.

We do not take a prescriptive stance on what these goals \textit{should} be.
We let the system decide what is reasonable.
These goals can, of course, also be the object of study in their own right; researchers can vary them or choose their own, but they are seemingly fundamental to any social science for reasons laid out in \cite{simon1996}.
Therefore, explicit goals are a requirement for agents in our system.

\subsection{Simulation design and execution}  \label{sec:interactions}

LLMs are designed to produce text.
And since an independent LLM powers each agent, one agent must finish speaking before the next begins.
So, in any multi-agent simulation, there must be a speaking order, which raises the question of how the system should determine this speaking order.
Unfortunately, most human conversations do not have an obvious order; people collectively figure out how to interact.
We centralize this process, but we could imagine a consensus protocol for who speaks next.

In more straightforward settings with only two agents (e.g., two people bargaining over a mug), the only possible conversational order is for the agents to alternate speaking.
As the number of agents in interaction increases beyond two, the number of possible speaking orders grows factorially.
For example, with three agents, there are $3! = 6$ ways to order them; with 4 agents, $4!=24$ orderings, and so on.
However, the number of possible orderings of the agents is only part of the complexity.

Who speaks next in a given conversation is a product of the participants' personalities, the setting of the conversation, the social dynamics between the speakers, the emotional state of the participants, and many other factors.
They are also adaptive---often, the speaking order changes throughout a conversation.
For example, in a court proceeding, the judge usually guides the interaction---signaling who speaks between the lawyers, witnesses, and the jury.
Each contributes at various and irregular intervals depending on both the type and stage of the proceeding.
In a family of two parents and two children, the order of who speaks next varies greatly.
It might depend on the parents' moods or how annoying the children have been that day.
In contrast, the teacher is typically the main speaker in a high school classroom, although this varies depending on the classroom activity, such as a lecture versus a group discussion. 
No simple universal formula exists for who speaks next in such diverse settings.

Like the aggregation methods for outcomes determined by multiple measurement questions, we designed a menu of six interaction protocols.
The system queries an LLM to select the appropriate protocol for a given scenario. 
Figure \ref{fig:interaction_types} provides the menu, and we discuss each in turn.

\subsubsection{Turn-taking protocols}

\begin{figure}[ht]
   \caption{Menu of interaction protocols for the system to choose from for a given scenario.}
   \label{fig:interaction_types}
    \centering
    \includegraphics[width=\linewidth]{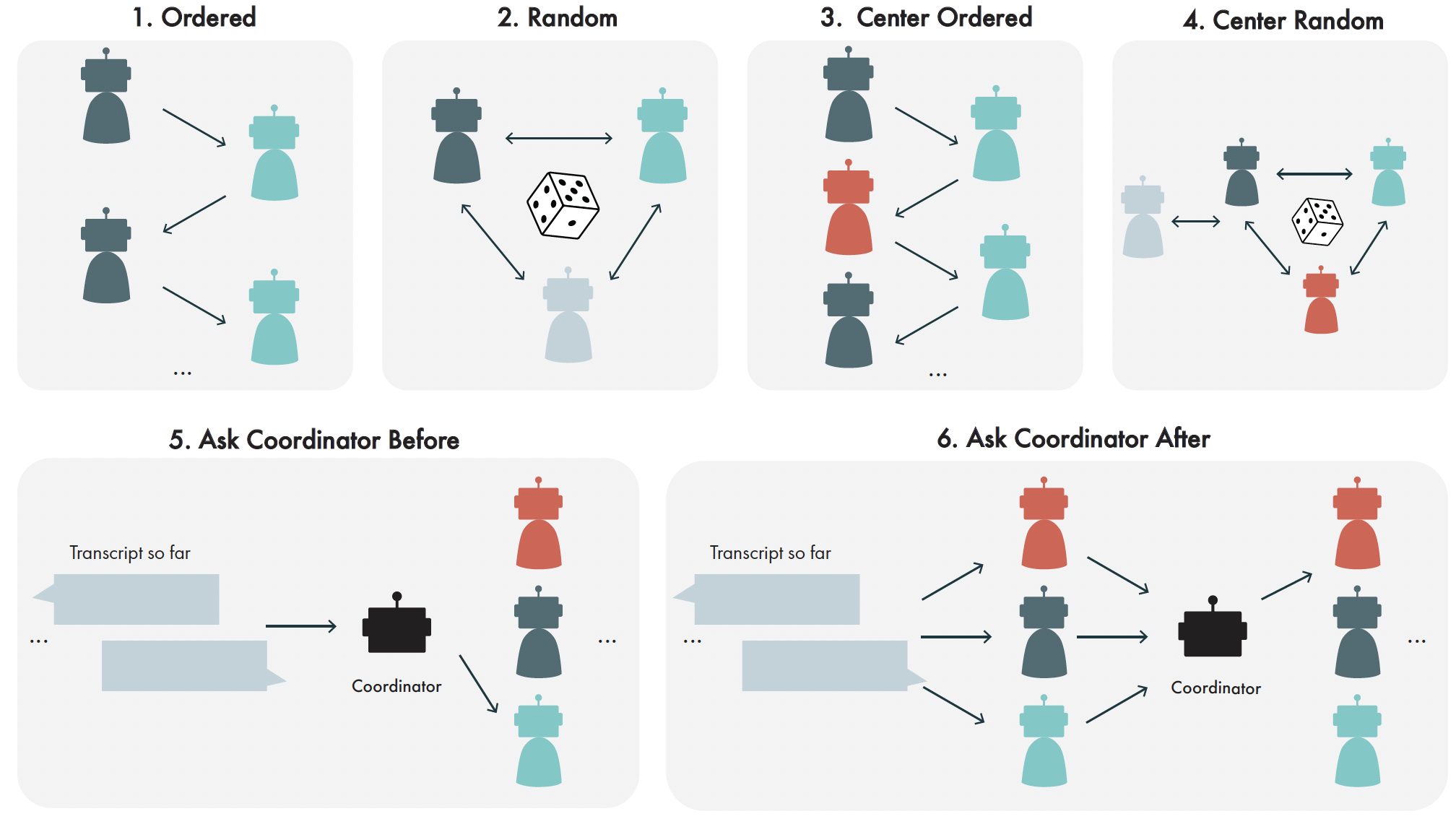}
    \begin{minipage}{\textwidth}
    \begin{footnotesize}
      \emph{Notes: (1) The agents speak in a predetermined order. (2) The agents speak in a random order. (3) A central agent alternates speaking with non-central agents in a predetermined order. (4) A central agent alternates speaking with non-central agents in random order. (5) A separate LLM (whom we call the coordinator) determines who speaks next based on the conversation. (6) Each agent responds privately to the conversation so far, and the coordinator realizes one of the responses.}
    \end{footnotesize}
    \end{minipage}
\end{figure}

The first interaction protocol is the \texttt{ordered} protocol (Figure \ref{fig:interaction_types}, option 1), where the agents speak in a predetermined order and continue repeatedly speaking in that order until the simulation is complete.
Next is the \texttt{random} protocol.
An agent is randomly selected to speak first (Figure \ref{fig:interaction_types}, option 2).
Then, each subsequent speaker is randomly selected, with the only restriction being that no agent can speak twice in a row.

In more complex scenarios with a central agent---an agent that speaks more than all others---like an auction with an auctioneer or a teacher in a classroom, the system can choose the \texttt{central-ordered} or \texttt{central-random} protocols (Figure \ref{fig:interaction_types}, options 3 and 4).
The former features a central agent who interacts alternately with a series of non-central agents, following a predetermined order among the non-central agents.
The latter also has a central agent alternating with the non-central agents but in random order.
Whenever there is an order of agents or a central agent, we also query the system to determine this order.

Finally, we designed two interaction protocols that provide more flexibility.
These interaction protocols involve a separate LLM-powered agent: ``the coordinator.''
The coordinator can read through transcripts of the conversations and make decisions about the simulations when necessary.
It can also answer measurement questions after the experiment.
The agents are not aware of the coordinator.
The use of the coordinator is the only part of the system that needs quasi-omniscient supervision.
Fortunately, LLMs perform so well that they can be used to automate this role.

In the \texttt{coordinator-before} protocol (Figure \ref{fig:interaction_types}, option 5), the coordinator is given the transcript of the conversation after each agent speaks. Then, it selects the next speaker.

In the \texttt{coordinator-after} protocol (Figure \ref{fig:interaction_types}, option 6), after each agent speaks, all the agents respond, but only the coordinator can see the responses along with the transcript of the conversation up to that point. 
Then, the coordinator chooses the response to ``realize'' as the real response.
The realized response is added to the conversation's transcript, and the rest are deleted as if they had never been made.
The only limitation in either of the coordinator protocols is that no agent can speak twice in a row.

\subsubsection{Executing the experimental simulations}

The system runs each experimental simulation in parallel, subject to the computational constraints of the researcher's machine. 
When the exogenous variable's values present too many combinations to sample from, a subset is randomly selected.
In every simulation, agents are provided with a description of the scenario, their unique private attributes, the other agents' roles, any public or scenario-level attributes, and access to the transcript of the conversation. 
Then, they interact according to the chosen interaction protocol.
However, none of the protocols specify when the simulation should end.

It is not obvious how to construct an optimal, nor even good, stopping rule.
Human conversations are unpredictable and do not always end when we expect them to or want them to \citep{conversations2021}.
An analogous issue is the halting problem in computer science, which is the problem of determining when, if ever, an arbitrary computer program will stop.
\cite{turing1936Halting} proved that no universal algorithm exists to solve the halting problem.

We implemented a two-tier mechanism to determine when to stop each simulation.
These apply to all interaction protocols.
After each agent speaks, the coordinator receives the transcript and decides if the conversation should continue---a yes or no decision.
Additionally, simulations are limited to 20 statements across all agents in the scenario, not including the coordinator.\footnote{
  Limiting the number of turns in the simulation is partially a convenience.
  As of the time of running the simulations for this paper, GPT-4 has a maximum token limit of 8,192 tokens, and the system must provide each agent with the entire conversation up to that point each time they need to speak.}
Agents are provided a live count of the remaining statements during the conversation.

\subsubsection{Post-simulation survey and data collection}  \label{sec:survey}

After the experiment, the system conducts a post-experiment survey.
As determined during the SCM construction, the system asks the relevant agents or the coordinator the survey questions to measure the outcome variable in each simulation.
The system then takes this question's raw answer and saves it as an observation along with the values of the exogenous variables.
If there is no reasonable answer to the question, say, if the outcome is conditional, then the system will report an \textit{NA} for the variable's value.

Once the system has the answer to the survey question, it queries an LLM with the survey question, the agent's response, and information about the variable's type to determine its correct numerical value as a string.
If the variable is a count or continuous variable, it is converted into an integer or a float.
If the variable is ordinal or binary, the system queries an LLM to map it to a whole-number integer sequence.
If multiple survey questions determine a variable, the system aggregates the answers to the questions using the method selected during the SCM construction phase.
Then, it converts the aggregated value to the appropriate type.
After parsing the data for each outcome, the system has a data frame with one column of numerical values for each variable in the SCM.

\subsection{Path estimation \& model fit}

With a complete dataset and the proposed SCM, the system can estimate the linear SCM without further queries to an LLM.
The system uses the R package lavaan to estimate all paths in the model \citep{lavaan2012}.\footnote{
  For those familiar with lavaan and Python, the system automatically generates the correctly formatted string in lavaan syntax using a Python dictionary that stores the structure of the SCM in key-value pairs.
  }
The system can standardize all estimates, estimate interactions and non-linear terms, and view various summary statistics for each variable.
It can also provide likelihood ratio, Wald, and Lagrange Multiplier tests to evaluate the model fit and compare path estimates.
The system can do any statistical estimation or test that is built into lavaan.

\subsection{Follow-on experiments}

Although we have not yet automated this process, the system can perform follow-on experiments.
Insignificant exogenous variables from the first experiment can be dropped.
Then, the system could query an LLM for new exogenous variables based on what might be interesting, given the already tested causal paths.
The system would use the same agents and interaction protocol, but the agents would vary on the new exogenous variables and the old ones that were significant in the first experiment.
Theoretically, the system can run follow-on experiments ad infinitum, and we can imagine future models that could be very good at proposing potential causal relationships.

\section{Hypotheses as structural causal models}  \label{sec:scms}

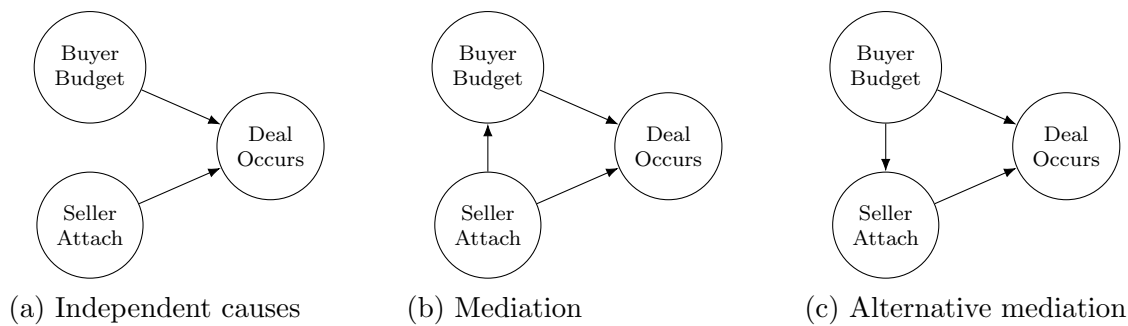
\begin{figure}
    \caption{Valid graphical interpretations of the same natural language hypothesis.}
    \label{fig:dag_moderation_examples}

    % Subfigure 1: No Moderation
    \begin{subfigure}{0.3\textwidth}
        \centering
        \begin{tikzpicture}[scale=0.6]
            % DAG nodes
            \node[draw, circle, align=center, font=\scriptsize] (budget) at (-1,0.5) {Buyer\\Budget};
            \node[draw, circle, align=center, font=\scriptsize] (attachment) at (-1,-3) {Seller \\ Attach};
            \node[draw, circle, align=center, font=\scriptsize] at (3,-1.25) (deal) {Deal\\ Occurs};
            % Arrows
            \draw[-Latex] (budget) -- (deal) ;
            \draw[-Latex] (attachment) -- (deal);
        \end{tikzpicture}
        \caption{Independent causes}
        \label{fig:dag_moderation_examples_1}
    \end{subfigure}
    \hfill
    \begin{subfigure}{0.3\textwidth}
        \centering
        \begin{tikzpicture}[scale=0.6]
            % DAG nodes
            \node[draw, circle, align=center, font=\scriptsize] (budget) at (-1,0.5) {Buyer\\Budget};
            \node[draw, circle, align=center, font=\scriptsize] (attachment) at (-1,-3) {Seller \\ Attach};
            \node[draw, circle, align=center, font=\scriptsize] at (3,-1.25) (deal) {Deal\\ Occurs};
            % Arrows
            \draw[-Latex] (budget) -- (deal) ;
            \draw[-Latex] (attachment) -- (deal) ;
            \draw[-Latex] (attachment) -- (budget) ;
        \end{tikzpicture}
        \caption{Mediation}
        \label{fig:dag_moderation_examples_2}
    \end{subfigure}
    \hfill
    % Subfigure 4: Full Moderation
    \begin{subfigure}{0.3\textwidth}
        \centering
        \begin{tikzpicture}[scale=0.6]
            % DAG nodes
            \node[draw, circle, align=center, font=\scriptsize] (budget) at (-1,0.5) {Buyer\\Budget};
            \node[draw, circle, align=center, font=\scriptsize] (attachment) at (-1,-3) {Seller \\ Attach};
            \node[draw, circle, align=center, font=\scriptsize] at (3,-1.25) (deal) {Deal\\ Occurs};
            % Arrows
            \draw[-Latex] (budget) -- (deal);
            \draw[-Latex] (attachment) -- (deal) ;
            \draw[-Latex] (budget) -- (attachment);
        \end{tikzpicture}
        \caption{Alternative mediation}
        \label{fig:dag_moderation_examples_3}
    \end{subfigure}
     \begin{minipage}{\textwidth}
    \begin{footnotesize}
      \emph{Notes: Each directed acyclic graph (DAG) is a valid causal interpretation of the following natural language hypothesis: ``The buyer's budget and the seller's sentimental attachment to the mug causally affect whether a deal occurs.''  
      In contrast, each DAG is unique in its declaration of the causal relationships.
      In DAGs, each arrow represents a direct causal relationship, and the absence of an arrow between two variables indicates no causal relationship. 
      If a variable is not included in the graph, then there is no stated causal relationship about this variable.
      While DAGs are unambiguous in their causal claims about which variables cause which other variables, they do not make any claims about the functional form of the relationships between variables.
      }
    \end{footnotesize}
    \end{minipage}
\end{figure}

Hypotheses stated in natural language can be ambiguous, making it challenging to discern precise implied causal relationships. 
Suppose a researcher is interested in two-person bargaining scenarios with a buyer and a seller.
And she has the following natural language hypothesis about two people bargaining over a mug: ``the buyer's budget and the seller's sentimental attachment to the mug causally affect whether a deal occurs.''
Figure \ref{fig:dag_moderation_examples} offers three ways we can interpret this causal statement: (\ref{fig:dag_moderation_examples_1}) the budget and the sentimental attachment could independently affect whether a deal occurs, (\ref{fig:dag_moderation_examples_2}) the budget could mediate the relationship between the attachment and the outcome, or (\ref{fig:dag_moderation_examples_3}), the mediation could be reversed.\footnote{
  This list of interpretations is not exhaustive.
}

For (\ref{fig:dag_moderation_examples_1}), an example could be an online marketplace where the buyer and seller cannot communicate.
When the buyer has a higher budget, she is more likely to buy the mug.
If the seller is more sentimentally attached to the mug, he may raise the price and, therefore, lower the probability of a deal.
However, without any form of communication, these causal variables would not affect each other.
For (\ref{fig:dag_moderation_examples_2}), if the buyer and the seller can communicate and the seller realizes that the buyer is willing to spend more, he might become more attached to the mug and value it higher because of the increased potential sale price.
Finally, for (\ref{fig:dag_moderation_examples_3}), the mediated relationship could be reversed.
If the buyer sees that the seller is attached to the mug, this may cause her to increase her budget, which increases the probability of a deal.
The ambiguity of stating even simple hypotheses makes natural language insufficient for our purposes.

The graphs in Figure \ref{fig:dag_moderation_examples} are directed acyclic graphs (DAGs) and represent causal relationships.
DAGs unambiguously state whether a variable is a direct cause of another variable---the direction of the arrow indicates the direction of the causal relationship \citep{Hernan2020WhatIf}.
The absence of an arrow between two variables indicates no causal relationship. 
If a variable is not included in the graph, then there is no stated causal relationship involving this variable.

While DAGs are clear in their claims about which variables cause others, they do not make any statements about the functional form of the relationships between variables.
In contrast, structural causal models unambiguously state the causal relationships between variables \emph{and} the functional forms of these relationships \citep{Pearl2016Primer}.

Structural causal models (SCM), as first explored by \cite{wright1934method}, represent hypotheses as sets of equations.
Suppose we assume the relationships between the variables in Figure \ref{fig:dag_moderation_examples} are linear.
We can write an SCM for each of the DAGs.
Figure~\ref{fig:dag_moderation_examples_1} can be stated as:
{
\setlength{\abovedisplayskip}{2pt}
\setlength{\belowdisplayskip}{2pt}
\begin{align}
  DealOccurs &= \beta_1 BuyerBudget + \beta_2 SellerAttachment  + \epsilon; \label{eq:deal_occurs_simple}
\end{align}

}
{
\setlength{\abovedisplayskip}{2pt}
\setlength{\belowdisplayskip}{2pt}
Figure~\ref{fig:dag_moderation_examples_2} as:
\begin{align}
  BuyerBudget &= \beta_0 SellerAttachment  + \eta \label{eq:seller_cause_buyer} \\
  DealOccurs &= \beta_1 BuyerBudget + \beta_2  SellerAttachment + \epsilon; \label{eq:buyer_mediate_seller}
\end{align}

}
and Figure~\ref{fig:dag_moderation_examples_3} as:
{
\setlength{\abovedisplayskip}{2pt}
\setlength{\belowdisplayskip}{2pt}
\begin{align}
  SellerAttachment &= \beta_0 BuyerBudget + \eta \label{eq:buyer_cause_seller} \\
  DealOccurs &= \beta_1 BuyerBudget + \beta_2 SellerAttachment + \epsilon. \label{eq:seller_mediate_buyer}
\end{align}

}

The set of equations that represent the causal relationships between variables make the SCM.
We could also write each SCM with interaction terms for some or all of the causes or even use other types of link functions, and these would all be equally valid representations of the corresponding DAGs.

\newpage \clearpage

\section{Additional figures and tables} \label{sec:appendix}

%interaction SCMs
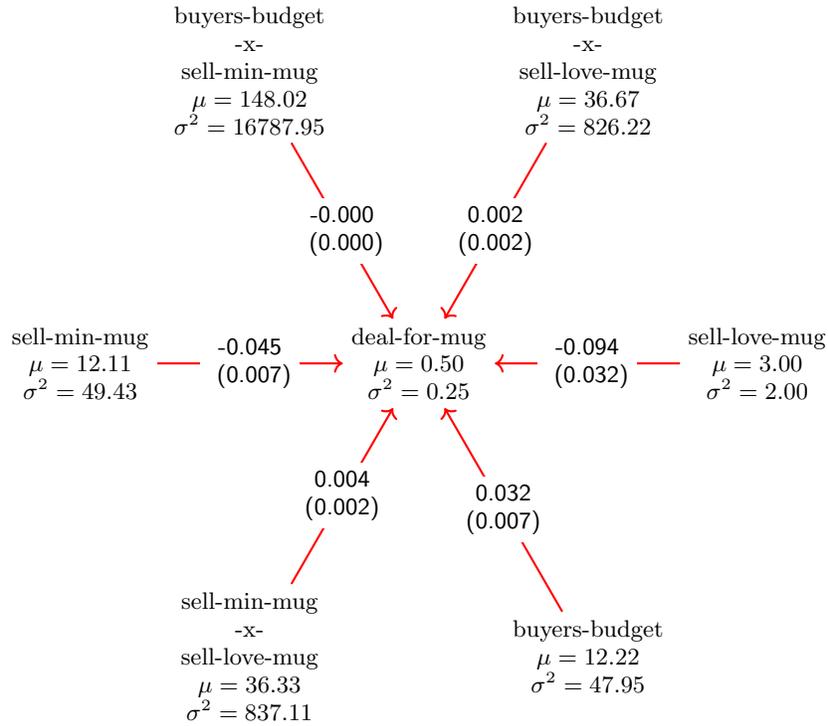
\begin{figure}[H]
\caption{Fitted SCM with interaction terms for ``two people bargaining over a mug."}
\label{fig:mug-love-interaction-scm}
\centering
\fontsize{9pt}{10.5pt}\selectfont

\begin{tikzpicture}
    \pgfdeclarelayer{background}
    \pgfdeclarelayer{foreground}
    \pgfsetlayers{background,main,foreground}
    \node[align=center] (deal-for-mug) at (0,0) {deal-for-mug\\ $\mu = 0.50$ \\ $\sigma^2 = 0.25$};
\node[align=center] (sell-love-mug) at (0:4.5cm) {sell-love-mug\\ $\mu = 3.00$ \\ $\sigma^2 = 2.00$};
\node[align=center] (buyers-budget-x-sell-love-mug) at (60.0:4.5cm) {buyers-budget\\-x-\\sell-love-mug\\ $\mu = 36.67$ \\ $\sigma^2 = 826.22$};
\node[align=center] (buyers-budget-x-sell-min-mug) at (120.0:4.5cm) {buyers-budget\\-x-\\sell-min-mug\\ $\mu = 148.02$ \\ $\sigma^2 = 16787.95$};
\node[align=center] (sell-min-mug) at (180.0:4.5cm) {sell-min-mug\\ $\mu = 12.11$ \\ $\sigma^2 = 49.43$};
\node[align=center] (sell-min-mug-x-sell-love-mug) at (240.0:4.5cm) {sell-min-mug\\-x-\\sell-love-mug\\ $\mu = 36.33$ \\ $\sigma^2 = 837.11$};
\node[align=center] (buyers-budget) at (300.0:4.5cm) {buyers-budget\\ $\mu = 12.22$ \\ $\sigma^2 = 47.95$};
\begin{pgfonlayer}{background}
\draw[->,red,thick] (buyers-budget) -- (deal-for-mug);
\draw[->,red,thick] (sell-min-mug) -- (deal-for-mug);
\draw[->,red,thick] (sell-love-mug) -- (deal-for-mug);
\draw[->,red,thick] (buyers-budget-x-sell-min-mug) -- (deal-for-mug);
\draw[->,red,thick] (buyers-budget-x-sell-love-mug) -- (deal-for-mug);
\draw[->,red,thick] (sell-min-mug-x-sell-love-mug) -- (deal-for-mug);
\end{pgfonlayer}
\begin{pgfonlayer}{foreground}
\path (buyers-budget) -- (deal-for-mug) node[midway, fill=white, font=\sffamily\fontsize{9pt}{10.5pt}\selectfont, align=center] {0.032\\(0.007)};
\path (sell-min-mug) -- (deal-for-mug) node[midway, fill=white, font=\sffamily\fontsize{9pt}{10.5pt}\selectfont, align=center] {-0.045\\\phantom{-}(0.007) };
\path (sell-love-mug) -- (deal-for-mug) node[midway, fill=white, font=\sffamily\fontsize{9pt}{10.5pt}\selectfont, align=center] {-0.094\\\phantom{-}(0.032) };
\path (buyers-budget-x-sell-min-mug) -- (deal-for-mug) node[midway, fill=white, font=\sffamily\fontsize{9pt}{10.5pt}\selectfont, align=center] {-0.000\\\phantom{-}(0.000) };
\path (buyers-budget-x-sell-love-mug) -- (deal-for-mug) node[midway, fill=white, font=\sffamily\fontsize{9pt}{10.5pt}\selectfont, align=center] {0.002\\(0.002)};
\path (sell-min-mug-x-sell-love-mug) -- (deal-for-mug) node[midway, fill=white, font=\sffamily\fontsize{9pt}{10.5pt}\selectfont, align=center] {0.004\\(0.002)};
\end{pgfonlayer}
\end{tikzpicture}\
  \begin{minipage}{\textwidth}
    \begin{footnotesize}
      \emph{\\ Notes: Each variable is given with its mean and variance. 
       The edges are labeled with their unstandardized path estimate and standard error. 
        There were 405 simulations with these agents: [`buyer', `seller'].}
    \end{footnotesize}
    \end{minipage}
\end{figure}
\newpage \clearpage
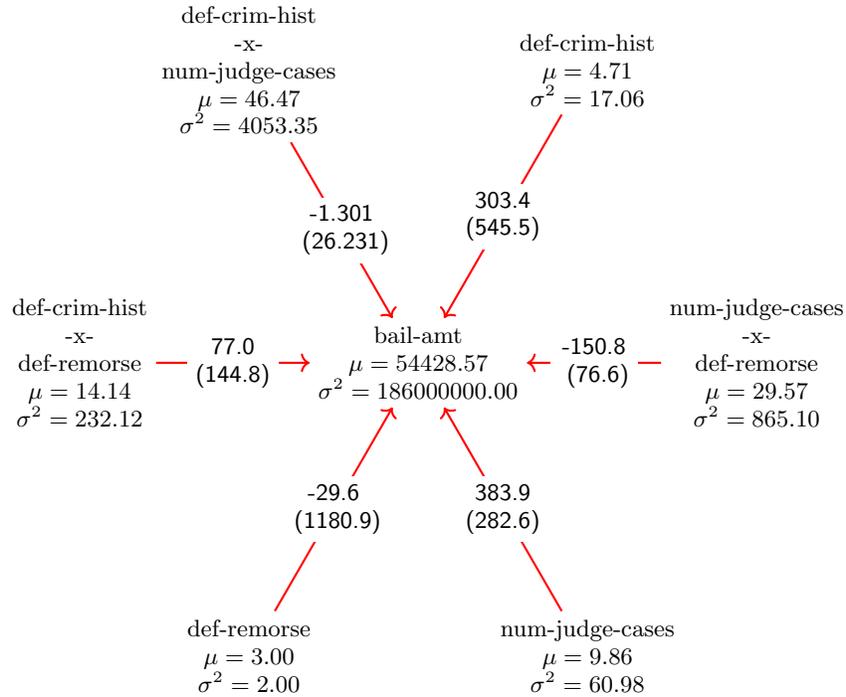
\begin{figure}[H]
\caption{Fitted SCM with interaction terms for ``a judge is setting bail for a criminal defendant who committed 50,000 dollars in tax fraud."}
\label{fig:tax-fraud-interaction-scm}
\centering
\fontsize{9pt}{10.5pt}\selectfont

\begin{tikzpicture}
    \pgfdeclarelayer{background}
    \pgfdeclarelayer{foreground}
    \pgfsetlayers{background,main,foreground}
    \node[align=center] (bail-amt) at (0,0) {bail-amt\\ $\mu = 54428.57$ \\ $\sigma^2 = 186000000.00$};
\node[align=center] (num-judge-cases-x-def-remorse) at (0:4.5cm) {num-judge-cases\\-x-\\def-remorse\\ $\mu = 29.57$ \\ $\sigma^2 = 865.10$};
\node[align=center] (def-crim-hist) at (60.0:4.5cm) {def-crim-hist\\ $\mu = 4.71$ \\ $\sigma^2 = 17.06$};
\node[align=center] (def-crim-hist-x-num-judge-cases) at (120.0:4.5cm) {def-crim-hist\\-x-\\num-judge-cases\\ $\mu = 46.47$ \\ $\sigma^2 = 4053.35$};
\node[align=center] (def-crim-hist-x-def-remorse) at (180.0:4.5cm) {def-crim-hist\\-x-\\def-remorse\\ $\mu = 14.14$ \\ $\sigma^2 = 232.12$};
\node[align=center] (def-remorse) at (240.0:4.5cm) {def-remorse\\ $\mu = 3.00$ \\ $\sigma^2 = 2.00$};
\node[align=center] (num-judge-cases) at (300.0:4.5cm) {num-judge-cases\\ $\mu = 9.86$ \\ $\sigma^2 = 60.98$};
\begin{pgfonlayer}{background}
\draw[->,red,thick] (def-crim-hist) -- (bail-amt);
\draw[->,red,thick] (num-judge-cases) -- (bail-amt);
\draw[->,red,thick] (def-remorse) -- (bail-amt);
\draw[->,red,thick] (def-crim-hist-x-num-judge-cases) -- (bail-amt);
\draw[->,red,thick] (def-crim-hist-x-def-remorse) -- (bail-amt);
\draw[->,red,thick] (num-judge-cases-x-def-remorse) -- (bail-amt);
\end{pgfonlayer}
\begin{pgfonlayer}{foreground}
\path (def-crim-hist) -- (bail-amt) node[midway, fill=white, font=\sffamily\fontsize{9pt}{10.5pt}\selectfont, align=center] {303.4\\(545.5)};
\path (num-judge-cases) -- (bail-amt) node[midway, fill=white, font=\sffamily\fontsize{9pt}{10.5pt}\selectfont, align=center] {383.9\\(282.6)};
\path (def-remorse) -- (bail-amt) node[midway, fill=white, font=\sffamily\fontsize{9pt}{10.5pt}\selectfont, align=center] {-29.6\\\phantom{-}(1180.9) };
\path (def-crim-hist-x-num-judge-cases) -- (bail-amt) node[midway, fill=white, font=\sffamily\fontsize{9pt}{10.5pt}\selectfont, align=center] {-1.301\\\phantom{-}(26.231) };
\path (def-crim-hist-x-def-remorse) -- (bail-amt) node[midway, fill=white, font=\sffamily\fontsize{9pt}{10.5pt}\selectfont, align=center] {77.0\\(144.8)};
\path (num-judge-cases-x-def-remorse) -- (bail-amt) node[midway, fill=white, font=\sffamily\fontsize{9pt}{10.5pt}\selectfont, align=center] {-150.8\\\phantom{-}(76.6) };
\end{pgfonlayer}
\end{tikzpicture}\
  \begin{minipage}{\textwidth}
    \begin{footnotesize}
      \emph{\\ Notes: Each variable is given with its mean and variance. 
       The edges are labeled with their unstandardized path estimate and standard error. 
        There were 245 simulations with these agents: [`judge', `defendant', `defense attorney', `prosecutor'].}
    \end{footnotesize}
    \end{minipage}
\end{figure}
\newpage \clearpage
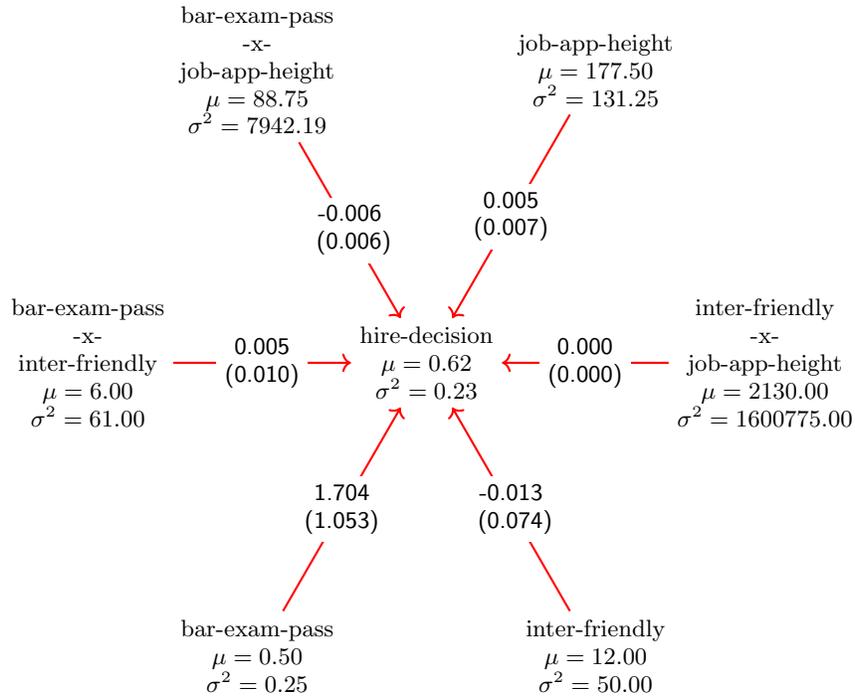
\begin{figure}[H]
\caption{Fitted SCM with interaction terms for ``a person is interviewing for a job as a lawyer."}
\label{fig:lawyer-interview-3var-interaction-scm}
\centering
\fontsize{9pt}{10.5pt}\selectfont

\begin{tikzpicture}
    \pgfdeclarelayer{background}
    \pgfdeclarelayer{foreground}
    \pgfsetlayers{background,main,foreground}
    \node[align=center] (hire-decision) at (0,0) {hire-decision\\ $\mu = 0.62$ \\ $\sigma^2 = 0.23$};
\node[align=center] (inter-friendly-x-job-app-height) at (0:4.5cm) {inter-friendly\\-x-\\job-app-height\\ $\mu = 2130.00$ \\ $\sigma^2 = 1600775.00$};
\node[align=center] (job-app-height) at (60.0:4.5cm) {job-app-height\\ $\mu = 177.50$ \\ $\sigma^2 = 131.25$};
\node[align=center] (bar-exam-pass-x-job-app-height) at (120.0:4.5cm) {bar-exam-pass\\-x-\\job-app-height\\ $\mu = 88.75$ \\ $\sigma^2 = 7942.19$};
\node[align=center] (bar-exam-pass-x-inter-friendly) at (180.0:4.5cm) {bar-exam-pass\\-x-\\inter-friendly\\ $\mu = 6.00$ \\ $\sigma^2 = 61.00$};
\node[align=center] (bar-exam-pass) at (240.0:4.5cm) {bar-exam-pass\\ $\mu = 0.50$ \\ $\sigma^2 = 0.25$};
\node[align=center] (inter-friendly) at (300.0:4.5cm) {inter-friendly\\ $\mu = 12.00$ \\ $\sigma^2 = 50.00$};
\begin{pgfonlayer}{background}
\draw[->,red,thick] (bar-exam-pass) -- (hire-decision);
\draw[->,red,thick] (inter-friendly) -- (hire-decision);
\draw[->,red,thick] (job-app-height) -- (hire-decision);
\draw[->,red,thick] (bar-exam-pass-x-inter-friendly) -- (hire-decision);
\draw[->,red,thick] (bar-exam-pass-x-job-app-height) -- (hire-decision);
\draw[->,red,thick] (inter-friendly-x-job-app-height) -- (hire-decision);
\end{pgfonlayer}
\begin{pgfonlayer}{foreground}
\path (bar-exam-pass) -- (hire-decision) node[midway, fill=white, font=\sffamily\fontsize{9pt}{10.5pt}\selectfont, align=center] {1.704\\(1.053)};
\path (inter-friendly) -- (hire-decision) node[midway, fill=white, font=\sffamily\fontsize{9pt}{10.5pt}\selectfont, align=center] {-0.013\\\phantom{-}(0.074) };
\path (job-app-height) -- (hire-decision) node[midway, fill=white, font=\sffamily\fontsize{9pt}{10.5pt}\selectfont, align=center] {0.005\\(0.007)};
\path (bar-exam-pass-x-inter-friendly) -- (hire-decision) node[midway, fill=white, font=\sffamily\fontsize{9pt}{10.5pt}\selectfont, align=center] {0.005\\(0.010)};
\path (bar-exam-pass-x-job-app-height) -- (hire-decision) node[midway, fill=white, font=\sffamily\fontsize{9pt}{10.5pt}\selectfont, align=center] {-0.006\\\phantom{-}(0.006) };
\path (inter-friendly-x-job-app-height) -- (hire-decision) node[midway, fill=white, font=\sffamily\fontsize{9pt}{10.5pt}\selectfont, align=center] {0.000\\(0.000)};
\end{pgfonlayer}
\end{tikzpicture}\
  \begin{minipage}{\textwidth}
    \begin{footnotesize}
      \emph{\\ Notes: Each variable is given with its mean and variance. 
       The edges are labeled with their unstandardized path estimate and standard error. 
        There were 80 simulations with these agents: [`job applicant', `employer'].}
    \end{footnotesize}
    \end{minipage}
\end{figure}
\newpage \clearpage
\begin{figure}[H]
\caption{Fitted SCM with interaction terms for ``3 bidders participating in an auction for a piece of art starting at fifty dollars."}
\label{fig:auction-art-3vars-interaction-scm}
\centering
\fontsize{9pt}{10.5pt}\selectfont

\begin{tikzpicture}
    \pgfdeclarelayer{background}
    \pgfdeclarelayer{foreground}
    \pgfsetlayers{background,main,foreground}
    \node[align=center] (final-art-price) at (0,0) {final-art-price\\ $\mu = 186.53$ \\ $\sigma^2 = 3867.92$};
\node[align=center] (bid1-max-budget-x-bid2-max-budg) at (0:4.5cm) {bid1-max-budget\\-x-\\bid2-max-budg\\ $\mu = 40000.00$ \\ $\sigma^2 = 900000000.00$};
\node[align=center] (bid1-max-budget) at (60.0:4.5cm) {bid1-max-budget\\ $\mu = 200.00$ \\ $\sigma^2 = 10000.00$};
\node[align=center] (bid2-max-budg-x-bid3-max-budg) at (120.0:4.5cm) {bid2-max-budg\\-x-\\bid3-max-budg\\ $\mu = 40000.00$ \\ $\sigma^2 = 900000000.00$};
\node[align=center] (bid3-max-budg) at (180.0:4.5cm) {bid3-max-budg\\ $\mu = 200.00$ \\ $\sigma^2 = 10000.00$};
\node[align=center] (bid1-max-budget-x-bid3-max-budg) at (240.0:4.5cm) {bid1-max-budget\\-x-\\bid3-max-budg\\ $\mu = 40000.00$ \\ $\sigma^2 = 900000000.00$};
\node[align=center] (bid2-max-budg) at (300.0:4.5cm) {bid2-max-budg\\ $\mu = 200.00$ \\ $\sigma^2 = 10000.00$};
\begin{pgfonlayer}{background}
\draw[->,red,thick] (bid1-max-budget) -- (final-art-price);
\draw[->,red,thick] (bid2-max-budg) -- (final-art-price);
\draw[->,red,thick] (bid3-max-budg) -- (final-art-price);
\draw[->,red,thick] (bid1-max-budget-x-bid2-max-budg) -- (final-art-price);
\draw[->,red,thick] (bid1-max-budget-x-bid3-max-budg) -- (final-art-price);
\draw[->,red,thick] (bid2-max-budg-x-bid3-max-budg) -- (final-art-price);
\end{pgfonlayer}
\begin{pgfonlayer}{foreground}
\path (bid1-max-budget) -- (final-art-price) node[midway, fill=white, font=\sffamily\fontsize{9pt}{10.5pt}\selectfont, align=center] {0.136\\(0.044)};
\path (bid2-max-budg) -- (final-art-price) node[midway, fill=white, font=\sffamily\fontsize{9pt}{10.5pt}\selectfont, align=center] {0.120\\(0.044)};
\path (bid3-max-budg) -- (final-art-price) node[midway, fill=white, font=\sffamily\fontsize{9pt}{10.5pt}\selectfont, align=center] {0.171\\(0.044)};
\path (bid1-max-budget-x-bid2-max-budg) -- (final-art-price) node[midway, fill=white, font=\sffamily\fontsize{9pt}{10.5pt}\selectfont, align=center] {0.001\\(0.000)};
\path (bid1-max-budget-x-bid3-max-budg) -- (final-art-price) node[midway, fill=white, font=\sffamily\fontsize{9pt}{10.5pt}\selectfont, align=center] {0.000\\(0.000)};
\path (bid2-max-budg-x-bid3-max-budg) -- (final-art-price) node[midway, fill=white, font=\sffamily\fontsize{9pt}{10.5pt}\selectfont, align=center] {0.000\\(0.000)};
\end{pgfonlayer}
\end{tikzpicture}\
  \begin{minipage}{\textwidth}
    \begin{footnotesize}
      \emph{\\ Notes: Each variable is given with its mean and variance. 
       The edges are labeled with their unstandardized path estimate and standard error. 
        There were 343 simulations with these agents: [`bidder 1', `bidder 2', `bidder 3', `auctioneer'].}
    \end{footnotesize}
    \end{minipage}
\end{figure}
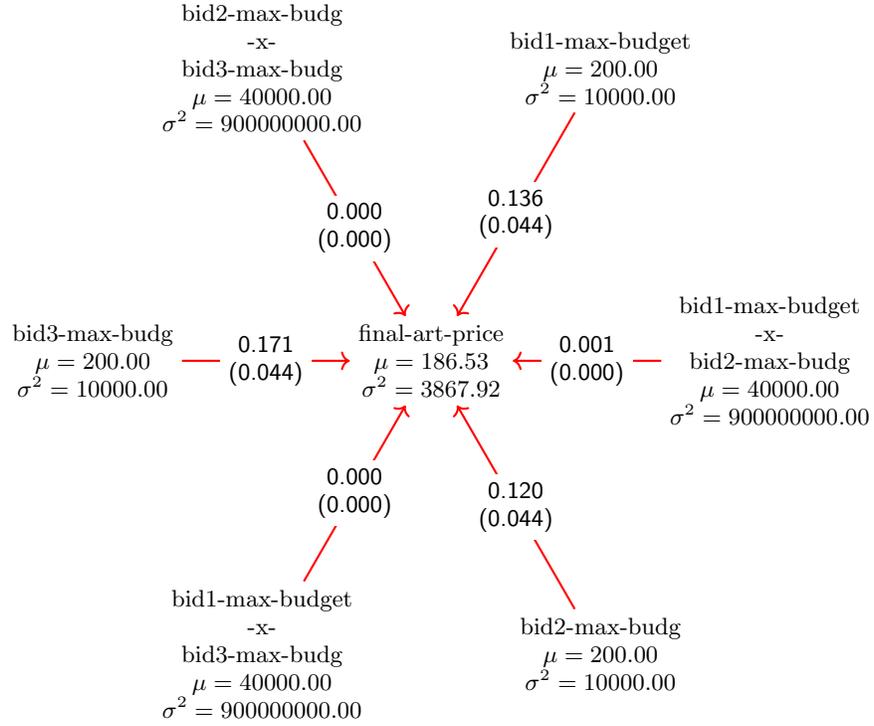

\begin{table}
  \caption{GPT-4's predictions for the path estimates for the experiments in Section~\ref{sec:results} at temperature 0.}
  \label{tab:predictions}
  \centering
  \scriptsize % Reducing font size
  \renewcommand{\arraystretch}{1.5}
  \begin{tabular}{
    |>{\centering\arraybackslash}p{1.75cm}|
    >{\centering\arraybackslash}p{1.75cm}|
    >{\centering\arraybackslash}p{1.75cm}|
    >{\centering\arraybackslash}p{1.25cm}|
    >{\centering\arraybackslash}p{1.5cm}|
    >{\centering\arraybackslash}p{1.25cm}|
  >{\centering\arraybackslash}p{1.5cm}|
    }
  \hline
  \textbf{Scenario (Outcome) } & \textbf{Exogenous Variable} & \textbf{Path Estimate} (SE) & \textbf{GPT-4 Guess} & \textbf{Two-tailed T-Test} & \textbf{GPT-4 Sign Correct} &$|\frac{\text{Predicted}}{\text{Experiment}}|$ Estimates\\ \hline
  \hline
  \multirow{3}{1.75cm}{\centering Mug Bargaining (Deal Made)} & Buyer's Budget & \BetaBudget\BudgetSig \space (\SEBudget)\phantom{{*}} & \BudgetPred\BudgetPredSig  &   $\BudgetTTestPval$ & \BudgetSign & \BudgetMag\\ \cline{2-7}
  & Seller's Min Price & \BetaMinPrice\MinPriceSig \space  (\SEMinPrice) & \MinPricePred\MinPricePredSig &  $\MinPriceTTestPval$ & \MinPriceSign & \MinPriceMag.00\\ \cline{2-7}
  & Seller's Attachment &\BetaLove\LoveSig \space (\SELove) & \LovePred\LovePredSig &  $\LoveTTestPval$ & \LoveSign & \LoveMag0\\ \hline
  \hline
  \multirow{3}{1.75cm}{\centering Art Auction (Final Price)} & Bidder 1 Budget &  \BetaBidOne\BidOneSig \space (\SEBidOne)\phantom{{*}} & \BidOnePred\BidOnePredSig & $\BidOneTestPval$ & \BidOneSign & \BidOneMag\\ \cline{2-7}
  & Bidder 2 Valuation & \BetaBidTwo\BidTwoSig \space (\SEBidTwo)\phantom{{*}} & \BidTwoPred\BidTwoPredSig & $\BidTwoTestPval$ & \BidTwoSign & \BidTwoMag\\ \cline{2-7}
  & Bidder 3 Valuation & \BetaBidThree\BidThreeSig \space (\SEBidThree\phantom{{*}}) &\BidThreePred\BidThreePredSig &$\BidThreeTestPval$ & \BidThreeSign & \BidThreeMag0\\ \hline
  \hline
  \multirow{3}{1.75cm}{\centering Bail Hearing (Bail Amount)} & Defendant's Previous Convictions & \BetaConvict\ConvictSig \space (\SEConvict)\phantom{{*}} & \ConvictPred\ConvictPredSig & $\ConvictTestPval$ & \ConvictSign & \ConvictMag\\ \cline{2-7}
  & Judge Cases That Day & \BetaCases\CasesSig \space (\SECases)\phantom{{*}} & \CasesPred\CasesPredSig & $\CasesTestPval$ & \CasesSign& \CasesMag\\ \cline{2-7}
  & Defendant's Remorse  & \BetaRemorse\RemorseSig\phantom{{*}} \space(\SERemorse)& \RemorsePred\RemorsePredSig & $\RemorseTestPval$ &\RemorseSign& \RemorseMag0\\ \hline
  \hline
  \multirow{3}{1.75cm}{\centering Lawyer Interview (Gets Job)} & Passed Bar & \BetaBar0\BarSig \space (\SEBar)\phantom{{*}} & \BarPred\BarPredSig & $\BarTestPval$ & \BarSign & \BarMag0\\ \cline{2-7}
  & Interviewer Friendliness & \BetaFriend \FriendSig \space (\SEFriend)&  \FriendPred \FriendPredSig & $\FriendTestPval$ & \FriendSign& \FriendMag.00 \\ \cline{2-7}
  & Applicant's Height & \BetaHeight \HeightSig \space (\SEHeight)\phantom{{*}}& \HeightPred\HeightPredSig & $\HeightTestPval$ & \HeightSign & \HeightMag\\ \hline
  \end{tabular}
  \begin{minipage}{\textwidth}
    \small
  \begin{footnotesize}
  \emph{Notes: 
  The table provides GPT-4's prediction for the path estimate for each experiment in Section~\ref{sec:results}
  From left to right, column 1 provides the scenario and outcome, column 2 provides the causal variable name, column 3 the path estimate and its standard error, and column 4 shows the LLM's prediction for the path estimate and whether it was predicted to be statistically significant. 
  Column 5 gives the p-value of a two-tailed t-test comparing the predictions to the results, column 6 is whether the predicted sign of the estimate was correct, and column 7 is the magnitude of the difference between the predicted and actual estimate.}
  \end{footnotesize}
  \end{minipage}
  \end{table}

\begin{table}
  \caption{GPT-4's predictions for the path estimates for the experiments in Section~\ref{sec:results} at temperature 1.}
  \label{tab:predictions_1}
  \centering
  \scriptsize % Reducing font size
  \renewcommand{\arraystretch}{1.5}
  \begin{tabular}{
    |>{\centering\arraybackslash}p{1.75cm}|
    >{\centering\arraybackslash}p{1.75cm}|
    >{\centering\arraybackslash}p{1.75cm}|
    >{\centering\arraybackslash}p{1.25cm}|
    >{\centering\arraybackslash}p{1.5cm}|
    >{\centering\arraybackslash}p{1.25cm}|
  >{\centering\arraybackslash}p{1.5cm}|
    }
  \hline
  \textbf{Scenario (Outcome) } & \textbf{Exogenous Variable} & \textbf{Path Estimate} (SE) & \textbf{GPT-4 Guess} & \textbf{Two-tailed T-Test} & \textbf{GPT-4 Sign Correct} (SE)&$|\frac{\text{Predicted}}{\text{Experiment}}|$ Estimates\\ \hline
  \hline
  \multirow{3}{1.75cm}{\centering Mug Bargaining (Deal Made)} & Buyer's Budget & \BetaBudget\BudgetSig \space (\SEBudget)\phantom{{*}} & \BudgetPredMean\BudgetPredSigMean \space (\BudgetPredSE)\phantom{{*}}&   $\BudgetTTestPvalMean$ & \BudgetSignMean & \BudgetMagMean\\ \cline{2-7}
  & Seller's Min Price & \BetaMinPrice\MinPriceSig \space  (\SEMinPrice) & \MinPricePredMean\MinPricePredSigMean \space (\MinPricePredSE)\phantom{{*}}&  $\MinPriceTTestPvalMean$ & \MinPriceSignMean & \MinPriceMagMean\\ \cline{2-7}
  & Seller's Attachment &\BetaLove\LoveSig \space (\SELove) & \LovePredMean\LovePredSigMean \space (\LovePredSE)\phantom{{*}}&  $\LoveTTestPvalMean$ & \LoveSignMean & \LoveMagMean\\ \hline
  \hline
  \multirow{3}{1.75cm}{\centering Art Auction (Final Price)} & Bidder 1 Budget &  \BetaBidOne\BidOneSig \space (\SEBidOne)\phantom{{*}} & \BidOnePredMean\BidOnePredSigMean \space (\BidOnePredSE)\phantom{{*}} & $\BidOneTestPvalMean$ & \BidOneSignMean & \BidOneMagMean\\ \cline{2-7}
  & Bidder 2 Valuation & \BetaBidTwo\BidTwoSig \space (\SEBidTwo)\phantom{{*}} & \BidTwoPredMean\BidTwoPredSigMean \space (\BidTwoPredSE)\phantom{{*}} & $\BidTwoTestPvalMean$ & \BidTwoSignMean & \BidTwoMagMean\\ \cline{2-7}
  & Bidder 3 Valuation & \BetaBidThree\BidThreeSig \space (\SEBidThree\phantom{{*}}) &\BidThreePredMean\BidThreePredSigMean \space (\BidThreePredSE)\phantom{{*}} &$\BidThreeTestPvalMean$ & \BidThreeSignMean & \BidThreeMagMean\\ \hline
  \hline
  \multirow{3}{1.75cm}{\centering Bail Hearing (Bail Amount)} & Defendant's Previous Convictions & \BetaConvict\ConvictSig \space (\SEConvict)\phantom{{*}} & \ConvictPredMean\ConvictPredSigMean \space (\ConvictPredSE)\phantom{{*}} & $\ConvictTestPvalMean$ & \ConvictSignMean & \ConvictMagMean\\ \cline{2-7}
  & Judge Cases That Day & \BetaCases\CasesSig \space (\SECases)\phantom{{*}} & \CasesPredMean\CasesPredSigMean \space (\CasesPredSE)\phantom{{*}} & $\CasesTestPvalMean$ & \CasesSignMean& \CasesMagMean\\ \cline{2-7}
  & Defendant's Remorse  & \BetaRemorse\RemorseSig\phantom{{*}} \space(\SERemorse)& \RemorsePredMean\RemorsePredSigMean \space (\RemorsePredSE00)& $\RemorseTestPvalMean$ &\RemorseSignMean& \RemorseMagMean\\ \hline
  \hline
  \multirow{3}{1.75cm}{\centering Lawyer Interview (Gets Job)} & Passed Bar & \BetaBar0\BarSig \space (\SEBar)\phantom{{*}} & \BarPredMean\BarPredSigMean \space (\BarPredSE)\phantom{{*}} & $\BarTestPvalMean$ & \BarSignMean & \BarMagMean\\ \cline{2-7}
  & Interviewer Friendliness & \BetaFriend \FriendSig \space (\SEFriend)&  \FriendPredMean \FriendPredSigMean \space (\FriendPredSE)\phantom{{*}} & $\FriendTestPvalMean$ & \FriendSignMean& \FriendMagMean \\ \cline{2-7}
  & Applicant's Height & \BetaHeight \HeightSig \space (\SEHeight)\phantom{{*}}& \HeightPredMean\HeightPredSigMean \space (\HeightPredSE)\phantom{{*}} & $\HeightTestPvalMean$ & \HeightSignMean & \HeightMagMean\\ \hline
  \end{tabular}
  \begin{minipage}{\textwidth}
    \small
  \begin{footnotesize}
  \emph{Notes: 
  The table provides GPT-4's prediction for the path estimate for each experiment in Section~\ref{sec:results}
  Each prediction is the average of 100 prompts at temperature 1.
  From left to right, column 1 provides the scenario and outcome, column 2 provides the causal variable name, column 3 the path estimate and its standard error, and column 4 shows the LLM's average prediction for the path estimate and whether it was predicted to be statistically significant more than 50\% of the time. 
  The given standard error is for the mean of the predictions, not the LLM's prediction for the standard error.
  Column 5 gives the p-value of a two-tailed t-test comparing the average prediction to the results, column 6 is whether the predicted sign of the estimate was correct more than 50\% of the time, and column 7 is the magnitude of the difference between the predicted and actual estimate.}
  \end{footnotesize}
  \end{minipage}
  \end{table}

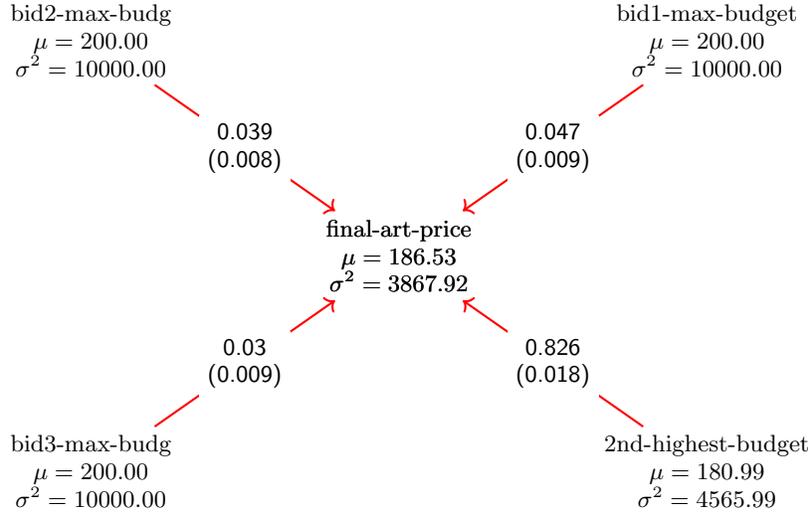
\begin{figure}[H]
  \caption{Fitted SCM for auction with bidder's reservation prices and second highest bid as exogenous variables.}
  \label{fig:auction-art-bids-theory}
  \centering
  \fontsize{9pt}{10.5pt}\selectfont
  
  \begin{tikzpicture}
      \pgfdeclarelayer{background}
      \pgfdeclarelayer{foreground}
      \pgfsetlayers{background,main,foreground}
      \node[align=center] (final-art-price) at (0,0) {final-art-price\\ $\mu = 186.53$ \\ $\sigma^2 = 3867.92$};
  \node[align=center] (bid1-max-budget) at (35.0:5cm) {bid1-max-budget\\ $\mu = 200.00$ \\ $\sigma^2 = 10000.00$};
  \node[align=center] (bid2-max-budg) at (145.0:5cm) {bid2-max-budg\\ $\mu = 200.00$ \\ $\sigma^2 = 10000.00$};
  \node[align=center] (bid3-max-budg) at (215.0:5cm) {bid3-max-budg\\ $\mu = 200.00$ \\ $\sigma^2 = 10000.00$};
  \node[align=center] (theory) at (325.0:5cm) {2nd-highest-budget\\ $\mu = \SCMTheorySimpleMean$ \\ $\sigma^2 = \SCMTheorySimpleVar$};
  \begin{pgfonlayer}{background}
  \draw[->,red,thick] (bid1-max-budget) -- (final-art-price);
  \draw[->,red,thick] (bid2-max-budg) -- (final-art-price);
  \draw[->,red,thick] (bid3-max-budg) -- (final-art-price);
  \draw[->,red,thick] (theory) -- (final-art-price);
  \end{pgfonlayer}
  \begin{pgfonlayer}{foreground}
  \path (bid1-max-budget) -- (final-art-price) node[midway, fill=white, font=\sffamily\fontsize{9pt}{10.5pt}\selectfont, align=center] {\SCMTheoryBidOneEst\\(\SCMTheoryBidOneSE)};
  \path (bid2-max-budg) -- (final-art-price) node[midway, fill=white, font=\sffamily\fontsize{9pt}{10.5pt}\selectfont, align=center] {\SCMTheoryBidTwoEst\\(\SCMTheoryBidTwoSE)};
  \path (bid3-max-budg) -- (final-art-price) node[midway, fill=white, font=\sffamily\fontsize{9pt}{10.5pt}\selectfont, align=center] {\SCMTheoryBidThreeEst\\(\SCMTheoryBidThreeSE)};
  \node[align=center] (final-art-price) at (0,0) {final-art-price\\ $\mu = 186.53$ \\ $\sigma^2 = 3867.92$};
  \path (theory) -- (final-art-price) node[midway, fill=white, font=\sffamily\fontsize{9pt}{10.5pt}\selectfont, align=center] {\SCMTheoryFullEst \\(\SCMTheoryFullSE)};

  \end{pgfonlayer}
  \end{tikzpicture}\
    \begin{minipage}{\textwidth}
      \begin{footnotesize}
        \emph{\\ Notes: Each variable is given with its mean and variance. 
         The edges are labeled with their unstandardized path estimate and standard error. 
          There were 343 simulations with these agents: [`bidder 1', `bidder 2', `bidder 3', `auctioneer'].}
      \end{footnotesize}
      \end{minipage}
  \end{figure}
  
\begin{figure}[H]
\caption{Fitted SCM for auction and second highest bid as exogenous variables.}
\label{fig:auction-art-theory}
\centering
\fontsize{9pt}{10.5pt}\selectfont
  \begin{tikzpicture}
    \pgfdeclarelayer{background}
    \pgfdeclarelayer{foreground}
    \pgfsetlayers{background,main,foreground}
    \node[align=center] (final-art-price) at (0,0) {final-art-price\\ $\mu = 186.53$ \\ $\sigma^2 = 3867.92$};
\node[align=center] (theory) at (180.0:6cm) {2nd-highest-budget\\ $\mu = \SCMTheorySimpleMean$ \\ $\sigma^2 = \SCMTheorySimpleVar$};
\begin{pgfonlayer}{background}
\draw[->,red,thick] (theory) -- (final-art-price);
\end{pgfonlayer}
\begin{pgfonlayer}{foreground}
\path (theory) -- (final-art-price) node[midway, fill=white, font=\sffamily\fontsize{9pt}{10.5pt}\selectfont, align=center] {\SCMTheorySimpleEst \\(\SCMTheorySimpleSE)};
\end{pgfonlayer}
\end{tikzpicture}\
  \begin{minipage}{\textwidth}
    \begin{footnotesize}
      \emph{\\ Notes: Each variable is given with its mean and variance. 
       The edges are labeled with their unstandardized path estimate and standard error. 
        There were 343 simulations with these agents: [`bidder 1', `bidder 2', `bidder 3', `auctioneer'].}
    \end{footnotesize}
    \end{minipage}
\end{figure}
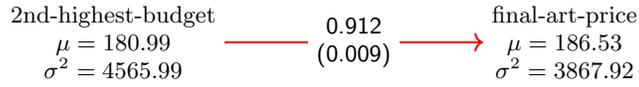

\begin{figure}[htbp!]
  \caption{Comparison of the LLM's predictions to the theoretical predictions and all experimental results for the auction scenario.}
  \label{fig:GPT-SCM-full}
  \centering
  \includegraphics[width=\linewidth]{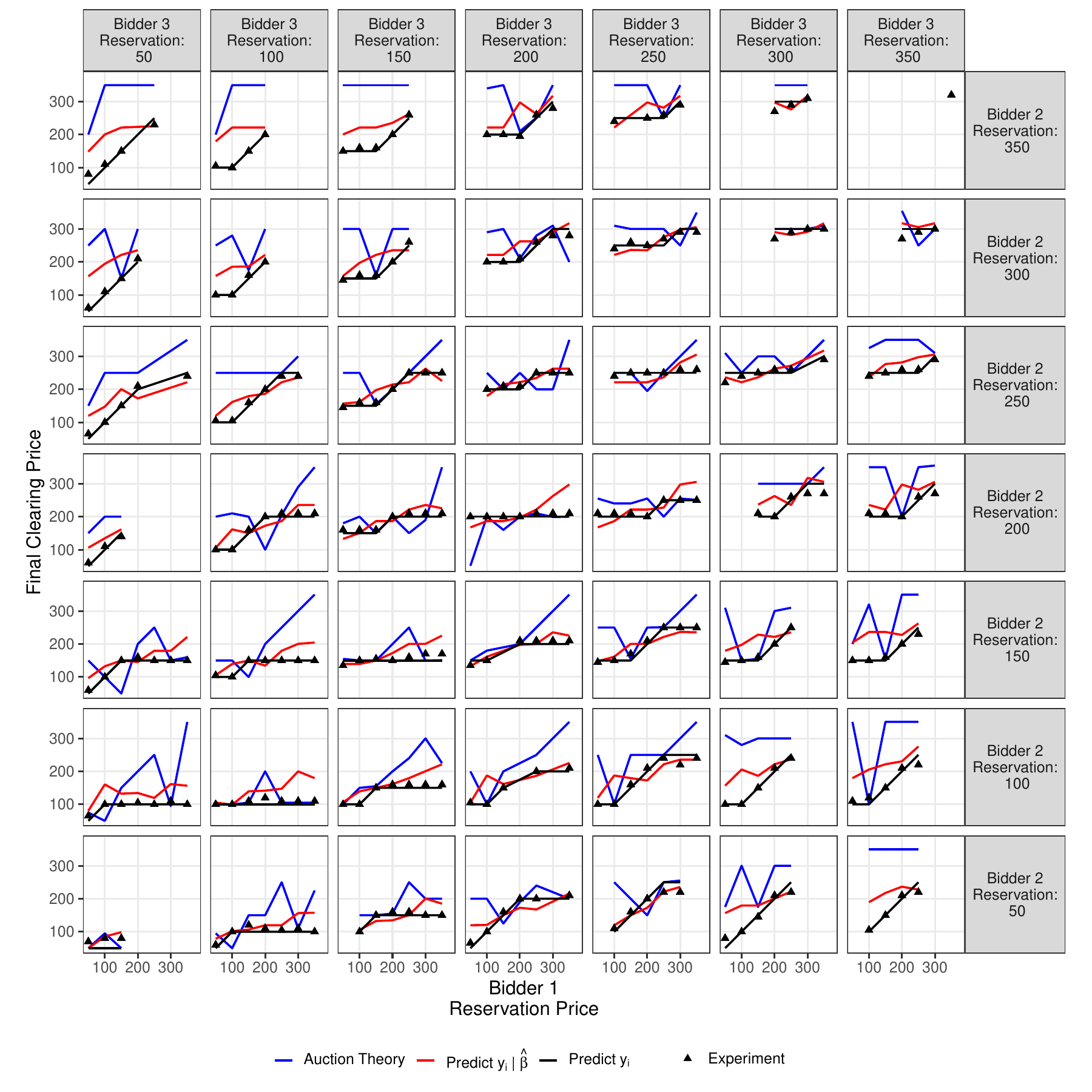}
  \begin{minipage}{\textwidth}
    \begin{footnotesize}
    \emph{Notes: The columns correspond to the different reservation values for bidder 3 in a given simulation, and the rows correspond to the different reservation values for bidder 2.
    The y-axis is the clearing price, and the x-axis lists bidder 1's reservation price.
    The black triangles track the observed clearing price in each simulated experiment, the black line shows the predictions made by auction theory ($MSE_{Theory} = \MSETheory$), the blue line indicates the LLM's predictions without the fitted SCM---the predict-$y_i$ task ($MSE_{y_i} = \MSEwoSCM$), and the red curve is the LLM's predictions with the fitted SCM---the predict-$y_i|\hat{\beta}_{-i}$ task ($MSE_{y_i|\hat{\beta}_{-i}} = \MSEwSCM$).
   }
    \end{footnotesize}
    \end{minipage}
\end{figure}

\newpage \clearpage

\begin{figure}[htbp!]
  \caption{Prompt used to elicity LLM predictions for the Predict-$\hat{\beta}$ task.}
  \label{fig:beta-predict}
  \centering
  
\begin{tcolorbox}[colback=blue!5!white,colframe=blue!75!black]
 
  I have just run an experiment to estimate the paths in the SCM from the TIKZ diagram below, which is delineated by triple backticks. We ran the experiment on multiple instances of GPT-4, once for each combination of the different ``Attribute Treatments'' in the accompanying table.
  This table also includes information about the variables and the individual agents involved in the scenario. 
  Your task is to predict the point estimates for the paths in the SCMs as accurately as possible based on the experiments. 
  You can see the summary statistics of the treatment variables below each variable name in the Tikz Diagram.
  We want to know how good you are at predicting the outcomes of experiments run on you. 
  Make sure you consider the correct units for both the cause and the outcome for each path.
  Please output your answer in the following form and do not include any other text:
  \{'predictions':  dictionary of point estimate predictions for each path\}
  \{'sig':  dictionary of whether or not each path is significant\} \texttt{```}Figure X and Table X\texttt{'''}
  \end{tcolorbox}
  
  \begin{minipage}{\textwidth}
    \begin{footnotesize}
    \emph{Notes: For each experiment, we input the accompanying table and the TIKZ diagram into the LLM between the triple backticks.
    For example, for the bargaining scenario, these are Figure~\ref{fig:mug-love-scm} and Table~\ref{tab:mug-love-table}.
   }
    \end{footnotesize}
    \end{minipage}
\end{figure}

\begin{table}[ht]
  \caption{Example of the information generated for each variable in an SCM.}
  \label{tab:variable_information}
  \centering
  \small
  \renewcommand{\arraystretch}{1.5}
  \begin{tabular}{
    |>{\centering\arraybackslash}p{3.6cm}|
    >{\centering\arraybackslash}p{2.8cm}|
    >{\centering\arraybackslash}p{3.1cm}|
    >{\centering\arraybackslash}p{3.8cm}|
  }
  \hline
  \textbf{Information Type} & \textbf{Deal Occurred} (Endogenous) & \textbf{Buyer's Budget} (Exogenous)& \textbf{Seller's Attachment} (Exogenous)\\
  \hline
  \textbf{Operationalization} & \texttt{1 if a deal occurs, 0 otherwise} & \texttt{Max amount the buyer will spend} & \texttt{Seller's emotional attachment level on a scale} \\
  \hline
  \textbf{Variable Type} & \texttt{Binary} & \texttt{Continuous} & \texttt{Ordinal} \\
  \hline
  \textbf{Units} & \texttt{Binary} & \texttt{Dollars} & \texttt{Levels of attachment} \\
  \hline
  \textbf{Levels} &  \texttt{\{0, 1\}} &  \texttt{\{\$0-\$5, \dots, \$40+\}} &  \texttt{\{\text{Low}, \dots, \text{High}\}} \\
  \hline
  \textbf{Explicit Measurement Questions} & \texttt{Buyer: ``Did a deal occur?''} & - & - \\
  \hline
  \textbf{Data Aggregation Method} & \texttt{Single Value} & - & - \\
  \hline
  \textbf{Scenario or Individual} & - & \texttt{Individual} & \texttt{Individual}\\
  \hline
  \textbf{Varied Attribute Proxies} & - &\texttt{``Your budget''} & \texttt{``Your attachment level''} \\
  \hline
  \textbf{Attribute Treatments} & - & \texttt{\{\$3, \dots, \$45\}} &  \texttt{\{\text{no attachment}, \dots, \text{extreme attachment}\}} \\
  \hline
  \end{tabular}
  \begin{minipage}{\textwidth}
    \begin{footnotesize}
    \emph{Notes: Each row shows a different piece of information generated for the variables in the SCM.
    The first column represents the type of information, the second column represents the information for the endogenous variable, and the third and fourth columns represent the information for the exogenous variables.
    This is example information based on the SCM in Figure \ref{fig:dag_moderation_examples_1}.}

    \end{footnotesize}
  \end{minipage}
\end{table}

\end{document}